\documentclass[twocolumn,showpacs,showkeys]{revtex4}
\usepackage{graphicx}
\usepackage{bm}
\usepackage{color}
\usepackage{amsmath}
\usepackage{natbib}

\begin{document}

\title{On equation of state for the "thermal" part of the spin current: Pauli principle contribution in the spin wave spectrum in cold fermion system}

\author{Pavel A. Andreev}
\email{andreevpa@physics.msu.ru}
\author{L. S. Kuz'menkov}%
\email{lsk@phys.msu.ru}
\affiliation{Faculty of physics, Lomonosov Moscow State University, Moscow, Russian Federation.}

 \date{\today}

\begin{abstract}
Spin evolution opened a large field in quantum plasma research. The spin waves in plasmas were considered among new phenomena considered in spin-1/2 quantum plasmas. The spin density evolution equation found by means of the many-particle quantum hydrodynamics shows existence of the "thermal" part of the spin current, which is an analog of the thermal pressure, or the Fermi pressure for degenerate electron gas, existing in the Euler equation. However, this term has been dropped, since there has not been found any equation of state for the thermal part of the spin current (TPSC), like we have for the pressure. In this paper we derive the equation of state for the TPSC and apply it for study of spectrum of collective excitations in spin-1/2 quantum plasmas. We focus our research on the spectrum of spin waves, since this spectrum is affected by the thermal part of the spin current. We consider two kinds of plasmas: electron-ion plasma with motionless ions and degenerate electrons, and degenerate electron-positron plasmas. We also present the non-linear Pauli equation with the spinor pressure term containing described effects. The thermal part of the flux of spin current existing in the spin current evolution equation is also derived. We also consider the contribution of the TPSC in the grand generalized vorticity evolution.
\end{abstract}


\maketitle




\section{\label{SCES sec:level1 Int} Introduction}

Almost 15 years the spin-1/2 quantum plasmas have been of the center of
active consideration. Starting from the explicit derivation of the continuity,
Euler, and the magnetic moment evolution (the Bloch equation) equations from
the many-particle Pauli equation in 2001 \cite{MaksimovTMP 2001},
\cite{MaksimovTMP 2001 b}, which contains the  derivation of the Coulomb
and spin-spin exchange correlations, analysis of many collective phenomena
has been performed.
Contribution of the spin dynamics (the magnetic moment dynamics) in the plasma
properties have been calculated during these research.
First of all the spin contribution in plasma dynamics arises via the force of the
spin-spin interaction existing in the Euler equation, requiring the account
of the Bloch equation.

On this path there have been studied the shift of the electromagnetic wave frequency \cite{Andreev VestnMSU 2007}, the increase of the fast magnetosonic mode frequency and the decrease of the slow magnetosonic
mode frequency \cite{Marklund PRL07}, \cite{Asenjo PL A 12}, the interaction of the magnetosonic waves in the spin-1/2 quantum plasmas \cite{Li and Han PP 14}, reduction of the energy transport in the quantum spin-1/2 plasmas due to the modification of the group velocity of the extraordinary wave at the certain range of wave numbers \cite{Li PP 12 122114}, widening of the solitary magnetosonic waves by a pressurelike term with negative sign caused by spin-spin interaction force \cite{Marklund PRE 07 067401}, non-linear whistlers in the strongly magnetized high density plasmas forming the large-scale density fluctuations \cite{Misra PRE 10 056406}, investigation of the magnetic diffusivity and obtaining that the magnetic diffusivity plays a
dominant role for the transition from the solitary wave to shock wave for arbitrary amplitude magnetosonic waves \cite{Sahu PP 13 112303}, detailed analysis of the small and arbitrary shock structures in spin 1/2 quantum plasma \cite{Sahu PP 15 022304}, composite nonlinear structures within the magnetosonic soliton interactions \cite{Han PP 15 062101}, the dynamics of small but finite amplitude magnetosonic waves exhibits both oscillatory and monotonic shock-like perturbations significantly affected by the spin-spin interaction \cite{Misra PLA 08 6412}, circularly polarized Alfven solitary waves with Gaussian form surrounded by smaller sinusoidal variations in the density envelope \cite{Keane PRE 11}, the modification of the Rayleigh–Taylor instability \cite{Modestov PP 09}.


The described phenomena, in some form, exist in spinless plasmas, while there
are some plasma effects requiring the spin of particles.
On this path the following purely spin plasma phenomena
have been found: the spin-plasma waves \cite{Andreev VestnMSU 2007}, \cite{Vagin 06}, \cite{Brodin PRL 08}, \cite{Misra JPP 10}, \cite{Andreev IJMP 12}, \cite{Zamanian PP 10},  \cite{Trukhanova PrETP 13}, \cite{Andreev PP 15 Positrons}, the spin-electron acoustic waves \cite{Andreev PRE 15 SEAW}, \cite{Andreev AoP 15 SEAW}, which are possibly related to the high-temperature superconductivity \cite{Andreev HTSC 15}, the spin-electron acoustic soliton \cite{Andreev 1504}, the spin (quantum) vorticity \cite{Andreev PP 15 Positrons}, \cite{Mahajan PRL 11}, \cite{Mahajan PP 15}, the spin caused modulational instability of the magnetosonic waves in the dense quantum plasma \cite{Misra PP 08 052105}, spin instabilities caused by specific equilibrium distribution functions \cite{Lundin PRE 10 056407}, instability of the plasma and spin-plasma waves at the propagation of the spin polarized neutron beam through the magnetized spin-1/2 plasmas arising due to the spin-spin and spin-current interactions Ref. \cite{Andreev IJMP 12}, \cite{Andreev AtPhys 08}. Some effects are reviewed in Ref. \cite{Shukla RMP 11}.

Moreover, the spin gives the contribution in the plasma dynamics via the modification of the Fermi pressure \cite{Andreev PRE 15 SEAW}, \cite{Andreev AoP 14 exchange} and due to the account of the spin current evolution \cite{Andreev IJMP B 15}.

The magnetization enters the spectrum of Langmuir waves propagating parallel to the external magnetic field due to the modification of the Fermi pressure \cite{Andreev PRE 15 SEAW}, \cite{Andreev AoP 14 exchange}. Increase of the spin polarization of electrons increases the Fermi pressure up to 60 percent. The contribution of this phenomenon can be also found in many other plasma effects.

The spin current tensor arises in the Bloch equation. Usually we present it  in terms of other hydrodymical variables, applying Takabayasi's results for single particle hydrodynamics \cite{Takabayasi PTP 55}, to truncate the set of quantum hydrodynamic (QHD) equations dropping its thermal part. However, we can find an equation for dynamical evolution of the spin current under influence of the electromagnetic field \cite{Andreev IJMP B 15}, \cite{Trukhanova PLA 15}, \cite{Brodin PPCF 11}, \cite{Stefan PRE 11}, \cite{Haas NJP 10}, but it does not give the thermal part of the spin current. In the general form the spin current evolution equation contains a pressure like term (the thermal part of the spin current flux), but no explicit form for this term has been found. Therefore this term has been dropped in earlier research. Let un mention now that in this paper we present the thermal part of the spin current (TPSC) and the thermal spin current flux for the degenerate electron gas, where the described quantities are caused by the Pauli blocking principle like the Fermi pressure in the Euler equation.

The spin current plays an important role in the quantum plasma physics. However, it is also in the center of attention in the condensed matter physics, especially in the application to spintronics. Spintronics is related to the spin-dependent electron transport phenomena \cite{Zutic RMP 04}. For the modeling of effects related to the spin currents the analytical definition of the spin current in terms of an effective single particle wave function is developed in Refs. \cite{Sun PRB 05}-\cite{Shi PRL 06}. Necessity of application of the full current composed of the spin and orbital currents is suggested in Ref. \cite{An Sci Rep 12}. Methodology of these research differs from the quantum hydrodynamics, where the spin current arises as a part of hydrodynamic variable set for the description of collective phenomena.

A special case of spin-1/2 quantum plasma modeling is the separate description of the spin-up and spin-down electrons. In this picture we do not consider electrons as a single fluid \cite{MaksimovTMP 2001}, \cite{Andreev VestnMSU 2007}, \cite{Brodin NJP 07}, but we consider the electrons as two interacting fluids \cite{Andreev PRE 15 SEAW}, \cite{Brodin PRL 10 SPF}, \cite{Harabadze RPJ 04}, \cite{Andreev 1410 Comment}. Since the Fermi pressures of the spin-up and spin-down magnetized electrons are different, the two fluid model gives extra longitudinal waves in the electron gas \cite{Andreev PRE 15 SEAW}, \cite{Andreev AoP 15 SEAW}. These waves are related to relative motion of the spin-up and spin-down electrons and called, due to their spectrum properties, the spin-electron acoustic waves (SEAWs). If we consider the wave propagation parallel or perpendicular to the external magnetic field we find a single SEAW \cite{Andreev PRE 15 SEAW}. While, at the oblique propagation, we find two branches of the SEAWs \cite{Andreev AoP 15 SEAW}. These modes are affected by the Coulomb exchange interaction \cite{Andreev 1504}. The nonlinear SEAWs propagating parallel to the external magnetic field were considered, the existence of the spin electron acoustic soliton was demonstrated \cite{Andreev 1504}. It was shown that interaction of electrons via the spelnons (the quantum of the SEAWs) gives a mechanism of the Cooper pairs formation \cite{Andreev HTSC 15}. This kind of Cooper pairs gives a mechanism of the high temperature superconductivity \cite{Andreev HTSC 15}. The thermal part of the spin current contributes in the separate spin evolution QHD either. Thus its derivation is essential for all forms of spin-1/2 QHDs.

QHD models are related to the specific form of the wave equation (Schrodinger, Pauli, or Dirac). In Ref. \cite{Ruiz arXiv 15} developed a general first-principle theory of resonant nondissipative vector waves assumes no specific wave equation.

Quantum properties of plasmas are also caused in the exchange interaction, but we do not consider it here. For the description and discussion of the exchange interaction in quantum plasmas see the following recent papers \cite{MaksimovTMP 2001 b}, \cite{Andreev 1504}, \cite{Andreev AoP 14 exchange}, \cite{Ekman 2015}, \cite{Zamanian PRE 13}, \cite{MaksimovTMP 1999}, \cite{Andreev PP 15 Cylinder}, \cite{Klimontovich JETP 73}.

Methodology of the many-particle quantum hydrodynamics can be applied to the derivation of the quantum kinetic equations for spinning particles \cite{Andreev Phys A 15}, which is an analog for the Wigner function method \cite{Wigner PR 84}.

Majority of effects in spin quantum plasma and spin related effects in condensed matter physics can be affected by the thermal part of the spin current. Thus, we obtain it here for the degenerate electron gas, so the thermal part of the spin current is related to the particle distribution under the Fermi step. It arises in the magnetic moment evolution equation. In its nature it does not related to the interparticle interaction. If one considers the spin-1/2 quantum plasmas with the spin-orbit interaction, when the spin current tensor arises in the force field in the Euler equation and the spin torque in the magnetic moment evolution equation \cite{Andreev IJMP 12}. In this regime the thermal part of the spin current contribution arises there either. Applying the thermal part of the spin current to different quantum plasma phenomena we consider the spin-plasmas waves propagating parallel and perpendicular to the external field in the electron-ion and the electron-positron plasmas, the thermal part of the spin current in the quantum vorticity, and calculate the thermal part of the spin current flux existing in the spin current evolution equation.

This paper is organized as follows. In Sec. II we present the basis of our model. We present the non-linear Pauli equation with the spinor pressure. We also describe the derivation of QHD equations from the microscopic model. In Sec. III we introduce the velocity field in the QHD equations and show the existence of the TPSC in the Bloch equation. In Sec. IV we present the explicit form of the thermal part of the spin current for the degenerate electron gas arising from the non-linear Pauli equation suggested in Sec. II. Corresponding modification of the quantum vorticity evolution equation is also described in Sec. IV. In Sec. V we present the thermal part of the spin current flux existing in the spin current evolution equation. This result is also based on the non-linear Pauli equation. In Sec. VI and below we apply the obtained model to the wave dispersion in plasmas. In Sec. VI we consider the contribution of the TPSC in spectrum of plasma waves, especially spin-plasma waves. In Sec. VII we show contribution of the thermal part of the spin current in the quantum hydrodynamic model of the electron-positron plasmas. In Sec. VIII we consider spectrum of the quantum electron-positron plasmas arising from presented model.
In Sec. IX a summary of obtained results is presented.

\section{\label{SCES sec:level1 Model} Model: QHD equations}

\subsection{\label{SCES sec:level1 NLSE} Non-linear Pauli equation}

Equations of the spin-1/2 quantum hydrodynamics can be represented as the Non-linear Pauli equation \cite{MaksimovTMP 2001}. It is similar to the spinless case when the continuity and Euler equations can be represented as the single fluid effective non-linear Schrodinger equation \cite{MaksimovTMP 1999}. Even spin-orbit interaction \cite{Andreev IJMP 12} and other relativistic effects \cite{Andreev PP 15 Positrons}, \cite{Mahajan IJTP 14} can be included in non-linear effective equations. Nonlinearity is caused by different factors: the many-particle interaction effect and the pressure, particularly the Fermi pressure.
The non-linear Schrodinger equation for two- and three-dimensional degenerate electron gases containing the Fermi pressure and the Coulomb exchange interaction for the partially polarized electron gas is obtained in Ref. \cite{Andreev AoP 14 exchange}. However, the partial spin polarization in the Fermi pressure presented in the scalar form. If we want to include the spin separation effect revealing in two fluid model of electrons \cite{Andreev PRE 15 SEAW}, \cite{Andreev AoP 15 SEAW}, \cite{Andreev 1410 Comment} we can construct the non-linear Pauli equation containing the pressure in the spinor form:
\begin{equation}\label{SCES Pressure spinor} \widehat{\pi}=\left(
                                                              \begin{array}{cc}
                                                                \pi_{\uparrow} & 0 \\
                                                                0 & \pi_{\downarrow} \\
                                                              \end{array}
                                                            \right),\end{equation}
which arises as a diagonal second rank spinor. Subindexes $\uparrow$ and $\downarrow$ refer to the spin-up and spin-down electrons. It can be represented in term of the Pauli matrixes $\widehat{\pi}=\pi_{\uparrow}(\hat{\textrm{I}}+\widehat{\sigma}_{z})/2+\pi_{\downarrow}(\hat{\textrm{I}}-\widehat{\sigma}_{z})/2$, where $\hat{\textrm{I}}$ is the unit second rank spinor $\hat{\textrm{I}}=\left(
\begin{array}{cc}
 1 & 0 \\
  0 & 1 \\
\end{array}
\right)$, and $\widehat{\sigma}_{z}$ is one of the Pauli matrixes
$\widehat{\sigma}_{z}=\left(
                        \begin{array}{cc}
                          1 & 0 \\
                          0 & -1 \\
 \end{array}\right)$.

In accordance with the Fermi pressures for the spin-up and spin-down electrons, $\pi_{s}$, with $s=\uparrow,$ $\downarrow$, has the following explicit form $\pi_{s}=(6\pi^{2})^{2/3}\hbar^{2}n_{s}^{2/3}/2m$. Here we have the mass of particles $m$, the reduced Planck constant $\hbar=1.06\times 10^{27}$ erg$\cdot$s, and the concentrations of spin-1/2 particles with the different spin projections on a chosen direction $n_{s}$.

In accordance with Refs. \cite{MaksimovTMP 2001}, \cite{Andreev IJMP 12}, \cite{Andreev PRE 15 SEAW} we can construct the non-linear Pauli equation with the Fermi pressure in the spinor form (\ref{SCES Pressure spinor}). We have
\begin{equation}\label{SCES Non Lin Pauli Eq} \imath\hbar\partial_{t}\Phi(\textbf{r},t)=\biggl(\frac{1}{2m}\widehat{\textbf{D}}^{2} +\widehat{\pi}+q\varphi -\gamma\widehat{\mbox{\boldmath $\sigma$}}\textbf{B}\biggr)\Phi(\textbf{r},t),\end{equation}
where $\widehat{\textbf{D}}=\widehat{\textbf{D}}(\textbf{r},t)=\widehat{\textbf{p}}-(q/c)\textbf{A}(\textbf{r},t)$, $q$ is the charge of particles, $c$ is the speed of light, $\varphi$ is the scalar potential of the electromagnetic field, $\textbf{B}$ is the magnetic field, $\textbf{A}$ is the vector potential of the electromagnetic field, hence $\textbf{B}=\nabla\times \textbf{A}$, $\gamma$ is the magnetic moment of particles, $\widehat{\mbox{\boldmath $\sigma$}}$ is the vector composed of the Pauli matrixes, $\Phi$ is the macroscopic spinor wave function related to the full concentration of particles as follows
$\Phi^{*}(\textbf{r},t)\Phi(\textbf{r},t)=n(\textbf{r},t)$.

We should mention that equation (\ref{SCES Non Lin Pauli Eq}) is actually in partial agreement with Ref. \cite{Andreev PRE 15 SEAW}. It corresponds to Ref. \cite{Andreev PRE 15 SEAW} in order of separate description of the spin-up and spin-down fermions and appearance of the partial Fermi pressures for each species of electrons. However equation (\ref{SCES Non Lin Pauli Eq}) leads to a thermal part of the spin current, which was not presented in Ref. \cite{Andreev PRE 15 SEAW}.

NLSE (\ref{SCES Non Lin Pauli Eq}) can be represented as the single fluid QHD of electrons or as the two fluid model with the separate spin evolution at our choice.

In the context of the model (\ref{SCES Pressure spinor}), (\ref{SCES Non Lin Pauli Eq}) we need to mention an excellent example of NLSEs applying at the study of Bose-Einstein condensates (BECs). It is well-known Gross-Pitaevskii equation. Its traditional form allows to model behavior of scalar particles (particles with zero spin or located in a single hyperfine state). However, the generalization of the Gross-Pitaevskii equation was suggested for the spinor BECs (the spin-1 and spin-2 bosons) with evolving occupations of three or five hyperfine states \cite{Ho PRL 98}-\cite{Stamper-Kurn RMP 13}. As it is shown in the mentioned papers, the evolution of the occupations leads to appearance of the spin waves. These waves are similar to the spin-plasma waves found for the magnetized spin-1/2 plasmas \cite{Andreev VestnMSU 2007}, \cite{Vagin 06}, \cite{Brodin PRL 08}. We should mention that in the neutral boson the nonlinearity in the NLSE is caused by the short range interaction \cite{Ho PRL 98}-\cite{Stamper-Kurn RMP 13}, while in plasmas it is related to the Fermi pressure and the electromagnetic interaction.

\subsection{\label{SCES sec:level1} First principles derivation of many-particle QHD equations}

The particle concentration is the quantum mechanical average of the operator constructed using the classic microscopic concentration
\begin{equation}\label{SCES concentration} n(\textbf{r},t)=\int \Psi^{+}(R,t)\sum_{i=1}^{N}\delta(\textbf{r}-\textbf{r}_{i})\Psi(R,t) dR,\end{equation}
where we integrate over the 3N dimensional configurational space, $dR=\prod_{i=1}^{N}d\textbf{r}_{i}$, and the sum of the Dirac delta functions is the operator of microscopic concentration.

Application of the microscopic concentration in classical physics allows to derive set of hydrodynamic-like equations \cite{Klimontovich book}, \cite{Weinberg book}. An explicit averaging of the microscopic concentration allows to obtain hydrodynamic equations for smooth functions describing macro behavior of mediums \cite{Kuz'menkov 91}, \cite{Andreev PIERS 2012 Cl}, \cite{Drofa TMP 96} (for more details see \cite{Kuz'menkov book}). Similarly, an appropriate definition of the quantum many particle concentration allows us to derive the set of quantum hydrodynamic equations with a truncation at necessary step.

To perform this derivation we need the explicit form of the Hamiltonian of the many-particle Schrodinger (Pauli) equation $\imath\hbar\partial_{t}\Psi=\hat{\textrm{H}}\Psi$. To include the major properties of the spin-1/2 electron-ion plasmas we include the Coulomb and spin-spin interaction
$$\hat{\textrm{H}}=\sum_{i}\biggl(\frac{1}{2m_{i}}\textbf{D}_{i}^{2}+q_{i}\varphi^{ext}_{i} -\gamma_{i}\widehat{\sigma}^{\alpha}_{i}\textrm{B}^{\alpha}_{i(ext)}\biggr) $$
\begin{equation}\label{SCES Hamiltonian gen}+\frac{1}{2}\sum_{i,j\neq i}(q_{i}q_{j}\textrm{G}_{ij}-\gamma_{i}\gamma_{j}\textrm{G}^{\alpha\beta}_{ij}\sigma^{\alpha}_{i}\sigma^{\beta}_{j}),\end{equation}
where $\textbf{D}_{i}=-\imath\hbar\nabla_{i}+q_{i}\textbf{A}_{i,ext}/c$.
Below we also consider the electron-positron plasmas. Features of spin-1/2 electron-positron plasmas obtained in Ref. \cite{Andreev PP 15 Positrons} will be briefly reviewed in the light of results of this paper. The complete model of spin-1/2 electron-positron plasmas will be applied to study the spectrum of the spin-plasma waves \cite{Andreev VestnMSU 2007}, \cite{Vagin 06}, \cite{Brodin PRL 08}, \cite{Andreev PP 15 Positrons}.

Let us describe meaning of different terms in Hamiltonian (\ref{SCES Hamiltonian gen}) and describe notations applied there. Hamiltonian (\ref{SCES Hamiltonian gen}) is composed of the kinetic energy operator, the potential energy of charges in the external electric field, the potential energy of the magnetic moments in the external magnetic field, the Coulomb interaction between charges, and the spin-spin interaction via the magnetic field created by the magnetic moments, correspondingly.
Subindexes $i$ and $j$ are the numbers of particles, $q_{i}$ and $m_{i}$ are the mass and charge of $i$th particle, $\gamma_{i}$ is the magnetic moment of $i$th particle, for electrons
$\gamma_{i}$ reads $\gamma_{e}=g_{e}q_{e}\hbar/(2m_{e}c)$, $q_{e}=-|e|$, and $g_{e}\approx1.00116$, difference of $g_{e}$ from the unit caused by the anomalous magnetic dipole moment.
The Green functions of the Coulomb, and the spin-spin interactions have the following form
$\textrm{G}_{ij}=1/r_{ij}$, $\textrm{G}^{\alpha\beta}_{ij}=4\pi\delta^{\alpha\beta}\delta(\textbf{r}_{ij})+\partial^{\alpha}_{i}\partial^{\beta}_{i}(1/r_{ij})$. We see that the Coulomb interaction is described by the scalar Green function, which is a solution of the Poisson equation $\nabla \textbf{E}=4\pi\rho$, where $\rho$ is the charge density. The Green function of the spin-spin interaction arises as a second rank symmetric tensor. It occurs as a solution of the following equations $\nabla\times \textbf{B}=4\pi\nabla\times \textbf{M}$ and $\nabla \textbf{B}=0$. In the Green function of the spin-spin interaction we apply the Kronecker symbol $\delta^{\alpha\beta}$ for the space components of vectors, which is the tensor representation of the unit matrix.
The quantities $\varphi^{ext}_{i}=\varphi(\textbf{r}_{i},t)$, $\textbf{A}_{i
(ext)}=\textbf{A}(\textbf{r}_{i},t)$ are the scalar and the vector
potentials of the external electromagnetic field:
$ \textbf{B}_{(ext)}(\textbf{r}_{i},t)=\nabla_{i}\times \textbf{A}_{(ext)}(\textbf{r}_{i},t),$ and
$\textbf{E}_{(ext)}(\textbf{r}_{i},t)=-\nabla_{i}\varphi_{ext}(\textbf{r}_{i},t) -\frac{1}{c}\partial_{t}\textbf{A}_{ext}(\textbf{r}_{i},t)$.
Operator $\widehat{\sigma}^{\alpha}_{i}$ are the Pauli matrixes, the commutation relations
for them are $ [\widehat{\sigma}^{\alpha}_{i},\widehat{\sigma}^{\beta}_{j}] =2\imath\delta_{ij}\varepsilon^{\alpha\beta\gamma}\widehat{\sigma}^{\gamma}_{i}$,
where we employ the Kronecker symbol on the particle numbers, which means that the commutator is equal to zero if we take the Pauli matrixes describing different particles. The commutator contains the Levi-Civita symbol $\varepsilon^{\alpha\beta\gamma}$, which is a third rank antisymmetric tensor.


The method under description describes multi species plasmas, such as electron-ion, electron-positron, or more complicate mixtures of species. However, for simplicity of presentation we focus our attention on a single species. We can do it since derivations of equations for different species are almost independent from each other. More explicit description of the derivation for electron-ion plasmas is described in Ref. \cite{Andreev Rev 14} (see formulae (28)-(45)).

At the first step, differentiating the particle concentration (\ref{SCES concentration}) with respect to time and applying the Schrodinger equation with the Hamiltonian (\ref{SCES Hamiltonian gen}) we find the continuity equation
\begin{equation}\label{SCES continuity equation}\partial_{t}n+\nabla \textbf{j}=0.\end{equation}

At derivation of the continuity equation (\ref{SCES continuity equation}) the explicit form of the particle current (the momentum density) appears as
\begin{equation}\label{SCES current definition} \textbf{j}=\int \sum_{i=1}^{N}\delta(\textbf{r}-\textbf{r}_{i})\frac{1}{2m_{i}}\biggl(\Psi^{+}(R,t)\textbf{D}_{i}\Psi(R,t)+h.c.\biggr) dR,\end{equation}
where h.c. means the hermitian conjugation.

Next we differentiate the particle current (\ref{SCES current definition}) with respect to time and find the momentum balance equation
\begin{equation} \label{SCES bal imp General structure} \partial_{t}\textrm{j}^{\alpha}+\frac{1}{m}\partial^{\beta}\Pi^{\alpha\beta}=\frac{1}{m}\textrm{F}^{\alpha},\end{equation}
where $\Pi^{\alpha\beta}$ is the momentum flux, and $\textbf{F}$ is the force field.

Force field consists of two parts
\begin{equation}\label{SCES Force sum} \textbf{F}=\textbf{F}_{ext}+\textbf{F}_{int}. \end{equation}

The explicit form of the momentum flux arises at the derivation of the Euler equation as follows
$$\Pi^{\alpha\beta}=\int \sum_{i=1}^{N}\delta(\textbf{r}-\textbf{r}_{i})\frac{1}{2m_{i}} \biggl(\Psi^{*}(R,t)\hat{\textrm{D}}_{i}^{\beta}\hat{\textrm{D}}_{i}^{\alpha}\Psi(R,t) $$
\begin{equation} +(\hat{\textrm{D}}_{i}^{\beta}\Psi)^{+}(R,t)\hat{\textrm{D}}_{i}^{\alpha}\Psi(R,t)+h.c.\biggr) dR.\end{equation}

The first term in formula (\ref{SCES Force sum}) is the force of the particle interaction with the external field
\begin{equation}\label{SCES Force ext} \textbf{F}_{ext}=qn\textbf{E}_{ext}+\frac{q}{c}\textbf{j}\times \textbf{B}_{ext}+\textrm{M}^{\beta}\nabla\textrm{B}^{\beta}_{ext}. \end{equation}
It consists of three terms describing the action the electric field on charges, the action of magnetic field on the moving charges, and the action of the external magnetic field on the magnetic moments, correspondingly.

The force field (\ref{SCES Force ext}) contains an extra function. In the last term we meet the magnetic moment density (the magnetization) $\textbf{M}$:
\begin{equation}\label{SCES magnetization}\textbf{M}(\textbf{r},t)=\int \sum_{i=1}^{N}\delta(\textbf{r}-\textbf{r}_{i})\gamma_{i}\Psi^{+}(R,t)\widehat{\mbox{\boldmath $\sigma$}}_{i}\Psi(R,t) dR.\end{equation}

The second part of the force field (\ref{SCES Force sum}) describes the Coulomb and spin-spin interparticle interactions
$$\textbf{F}_{int}=-q^{2}\int (\nabla \textrm{G}(\textbf{r}-\textbf{r}')) n_{2}(\textbf{r},\textbf{r}',t) d\textbf{r}' $$
\begin{equation}\label{SCES Force int}
+\int (\nabla \textrm{G}^{\beta\gamma}(\textbf{r}-\textbf{r}')) \textrm{M}_{2}^{\beta\gamma}(\textbf{r},\textbf{r}',t) d\textbf{r}', \end{equation}
where
\begin{equation}\label{SCES two part conc} n_{2}(\textbf{r},\textbf{r}',t)=\int \sum_{i,j\neq i} \delta(\textbf{r}-\textbf{r}_{i})\delta(\textbf{r}-\textbf{r}'_{j}) \Psi^{*}(R,t)\Psi(R,t) dR\end{equation}
is the two-particle concentration, and
$$\textrm{M}_{2}^{\alpha\beta}(\textbf{r},\textbf{r}',t)=\int \sum_{i,j\neq i} \delta(\textbf{r}-\textbf{r}_{i})\delta(\textbf{r}-\textbf{r}'_{j})\times$$
\begin{equation}\label{SCES two part magn}  \times\gamma_{i}\gamma_{j}\Psi^{*}(R,t)\sigma^{\alpha}_{i}\sigma^{\beta}_{j}\Psi(R,t) dR\end{equation}
is the two-particle magnetization.

We have presented a general derivation of the Euler equation from the many-particle Schrodinger equation. Therefore, the force field (\ref{SCES Force int}) arose beyond the self-consistent field approximation.
Extracting the self-consistent field we can present a general two particle function in the following form
$f_{2}(\textbf{r},\textbf{r}',t)=f(\textbf{r},t)f(\textbf{r}',t)+g_{2}(\textbf{r},\textbf{r}',t)$, where we have introduced the correlation function $g_{2}(\textbf{r},\textbf{r}',t)$.
In this paper we restrict our analysis by the self-consistent field approximation. Hence we drop the correlations. We consider the two particle concentration as the product of the particle concentrations
$n_{2}(\textbf{r},\textbf{r}',t)=n(\textbf{r},t)n(\textbf{r}',t)$.
Similarly, for the two-particle magnetization we present it as the product of the
magnetization
$\textrm{M}^{\alpha\beta}_{2}(\textbf{r},\textbf{r}',t)=\textrm{M}^{\alpha}(\textbf{r},t)\textrm{M}^{\beta}(\textbf{r}',t)$.
Account of the exchange interaction leads us beyond the self-consistent field approximation. Corresponding references are presented in the introduction. Let us mention that most of sited particles are focused on the Coulomb exchange interaction. The spin-spin exchange interaction is considered in Ref. \cite{MaksimovTMP 2001 b}.

The self-consistent electric and magnetic fields appear in the following form
\begin{equation} \label{SCES El Field} \textbf{E}=-q\nabla\int  \textrm{G}(\textbf{r},\textbf{r}') n(\textbf{r}',t) d\textbf{r}',\end{equation}
and
\begin{equation} \label{SCES Magn Field} \textrm{B}^{\alpha}=\int  \textrm{G}^{\alpha\beta}(\textbf{r},\textbf{r}') \textrm{M}^{\beta}(\textbf{r}',t) d\textbf{r}',\end{equation}
or, in more explicit, vector, form we have
\begin{equation}\textbf{B}=\nabla\int\biggl[\biggl(\nabla\frac{1}{\mid \textbf{r}-\textbf{r}'\mid}\biggr)\textbf{M}(\textbf{r}'.t)\biggr]d\textbf{r}'+4\pi \textbf{M}. \end{equation}

The introduced electric (\ref{SCES El Field}) and magnetic (\ref{SCES Magn Field}) fields satisfy the quasi static Maxwell equations $\nabla \textbf{E}=4\pi\rho$, $\nabla \times\textbf{E}=0$, $\nabla \times\textbf{B}=4\pi\nabla \times\textbf{M}$, $\nabla \textbf{B}=0$, where $\rho$ is the charge density.

In the self-consistent field approximation the full force field can be presented as follows
\begin{equation}\label{SCES Force SCF 1} \textbf{F}=qn\textbf{E}+\frac{q}{c}n\textbf{v}\times\textbf{B}+\textrm{M}^{\beta}\nabla \textrm{B}^{\beta}. \end{equation}

Considering the time evolution of the magnetic moment density defined by formula (\ref{SCES magnetization}), and applying the Schrodinger equation with Hamiltonian (\ref{SCES Hamiltonian gen}), we find, in the self-consistent field approximation, the magnetic moment evolution equation
\begin{equation}\label{SCES magn mom balance eq}\partial_{t}\textrm{M}^{\alpha}+\nabla^{\beta}\textrm{J}^{\alpha\beta} =\frac{2\gamma}{\hbar}\varepsilon^{\alpha\beta\gamma}\textrm{M}^{\beta}\textrm{B}^{\gamma}\end{equation}
containing the spin current tensor
$$\textrm{J}^{\alpha\beta}=\int \sum_{i=1}^{N}\delta(\textbf{r}-\textbf{r}_{i})\frac{\gamma_{i}}{2m_{i}}\times$$
\begin{equation}\label{SCES spin current}\times\biggl(\Psi^{+}(R,t)\hat{\textrm{D}}_{i}^{\beta}\widehat{\sigma}_{i}^{\alpha}\Psi(R,t)+h.c.\biggr) dR. \end{equation}
General form of the magnetic moment evolution equation can be found in Ref. \cite{MaksimovTMP 2001} (see formula 30).

\section{\label{SCES sec:level1} Introduction of the velocity field}

In traditional classic and quantum hydrodynamic equations we do not meet the particle current (the momentum density) $\textbf{j}$, the momentum flux tensor $\Pi^{\alpha\beta}$, and the spin current $\textrm{J}^{\alpha\beta}$. Instead we meet the velocity field $\textbf{v}$, thermal pressure tensor $\textrm{P}^{\alpha\beta}$, and expect to see the thermal part of the spin current $\textrm{J}^{\alpha\beta}_{th}$, which has been assumed to be equal to zero. Therefore, it is our job to separate the center mass motion and the thermal motion. This problem was considered, for instance, in Refs. \cite{MaksimovTMP 2001}, \cite{Trukhanova PrETP 13}, \cite{Andreev Asenjo}. In spinless case it was effectively addressed by the Madelung decomposition \cite{MaksimovTMP 1999} applied for the many-particle wave function.
For the spin-1/2 particles this problem requires a generalization of the Madelung decomposition, see for instance \cite{Takabayasi PTP 55}, presented by formula
\begin{equation}\label{SCES Madelung generilized} \psi(\textbf{r},t)=a(\textbf{r},t)e^{\imath S(\textbf{r},t)/\hbar}\phi(\textbf{r},t),\end{equation}
which contains the unit spinor $\phi(\textbf{r},t)$. Its explicit form is
\begin{equation}\label{SCES} \phi(\textbf{r},t)=\left(
                                                     \begin{array}{c}
                                                       \cos(\theta/2) e^{\imath \varphi/2} \\
                                                       \imath \sin(\theta/2) e^{-\imath \varphi/2} \\
                                                     \end{array}
                                                   \right).\end{equation}
This spinor satisfy the next condition $\phi^{+}\phi=1$. The generalization Madelung decomposition contains $a$ and $S$, which are the amplitude and phase of wave function correspondingly.  Same representation can be applied to the macroscopic wave function obeying the NLSE (\ref{SCES Non Lin Pauli Eq}) and microscopic many-particle wave function obeying the Schrodinger equation with Hamiltonian (\ref{SCES Hamiltonian gen}).

Since our analysis is focused on the spin current let us present the explicit form for the spin current for the single particle in the external field
\begin{equation}\label{SCES spin current one part} \textrm{J}^{\alpha\beta}_{sp}=\gamma ns^{\alpha}v^{\beta}-\frac{\hbar\gamma}{2m}\varepsilon^{\alpha\mu\nu}\biggl(ns^{\mu}\partial^{\beta}s^{\nu}\biggr)\end{equation}
arising from the generalized Madelung decomposition, where $n(\textbf{r},t)=$$\mid\psi(\textbf{r},t)\mid=a^{2}$ is the single-particle concentration, $\textbf{s}=\phi^{+}\widehat{\mbox{\boldmath $\sigma$}}\phi$ is the single-particle spin field, $\widehat{\mbox{\boldmath $\sigma$}}$ are the Pauli matrixes, $\textbf{v}=\nabla S/m-(\imath\hbar/m)\phi^{+}\nabla\phi-(q/mc)\textbf{A}$ is the single particle velocity field. The first term is the kinetic (classic-like) part of the spin current and the second term is the quantum part of the spin current, which is an analog of the quantum Bohm potential being the quantum part of the pressure.

At the analysis of many particle behavior we obtained more general formula, which contains an extra term
\begin{equation}\label{SCES spin current many part} \textrm{J}^{\alpha\beta}=\textrm{M}^{\alpha}v^{\beta} -\frac{\hbar}{2m\gamma}\varepsilon^{\alpha\mu\nu}\textrm{M}^{\mu}\partial^{\beta}\Biggl(\frac{\textrm{M}^{\nu}}{n}\Biggr)+\textrm{J}^{\alpha\beta}_{th}.\end{equation}
This term is related to incoherent motion of spinning particles and presents the thermal part of the spin current.

In the many particle case formula $\textbf{v}_{i}(R,t)=\nabla_{i} S(R,t)/m_{i}-(\imath\hbar/m_{i})\phi^{+}(R,t)\nabla\phi(R,t)-(q_{i}/m_{i}c)\textbf{A}_{i}$ describes quantum velocity of $i$th particle in the system. The velocity field is defined as $\textbf{v}=\textbf{j}/n$. The difference between the particle velocities and the local center of mass velocity (the velocity field) gives us the thermal velocity of particles. Similar description can be given for the collective part and thermal part of the spin density, for details see Ref. \cite{Andreev Asenjo}.

Application of the generalized Madelung decomposition to the momentum flux tensor $\Pi^{\alpha\beta}$ in the many particle Euler equation gives the following result $\Pi^{\alpha\beta}=mnv^{\alpha}v^{\beta}+\textrm{P}^{\alpha\beta}+\textrm{T}^{\alpha\beta}$, where $\textrm{P}^{\alpha\beta}$ is the thermal pressure, and $\textrm{T}^{\alpha\beta}$ is the quantum part giving the quantum Bohm potential. Similar picture arises for the spin current $\textrm{J}^{\alpha\beta}=n\mu^{\alpha} v^{\beta}+\textrm{J}^{\alpha\beta}_{quant}+\textrm{J}^{\alpha\beta}_{th}$, where $\textbf{M}=n\mbox{\boldmath $\mu$}$, $\textrm{J}^{\alpha\beta}_{quant}$ is the quantum part of the spin current \cite{Takabayasi PTP 55}, \cite{Andreev Asenjo}, and $\textrm{J}^{\alpha\beta}_{th}$ is the thermal part of the spin current \cite{MaksimovTMP 2001}, \cite{Andreev Asenjo}. Substituting the particle current $\textbf{j}=n\textbf{v}$, the momentum flux $\Pi^{\alpha\beta}$, and the spin current $\textrm{J}^{\alpha\beta}$ in the continuity, Euler and magnetic moment evolution equations we obtain for them the following representation:
\begin{equation}\label{SCES continuity equation with Velocity} \partial_{t}n+\nabla (n\textbf{v})=0,\end{equation}
$$mn(\partial_{t}+\textbf{v}\nabla)\textbf{v}+\nabla\textrm{P}-\frac{\hbar^{2}}{4m}n\nabla\Biggl(\frac{\triangle n}{n}-\frac{(\nabla n)^{2}}{2n^{2}}\Biggr)$$
\begin{equation}\label{SCES momentum balance eq with Velocity} +\frac{\hbar^{2}}{4m\gamma^{2}}\partial^{\beta}\biggl(n(\partial^{\beta}\mu^{\gamma})\nabla\mu^{\gamma}\biggr)=qn\textbf{E}+\frac{q}{c}n\textbf{v}\times \textbf{B}+\textrm{M}^{\beta}\nabla\textrm{B}^{\beta},
\end{equation}
and 
\begin{equation}\label{SCES magnetric moment evol Eq with velocity} n(\partial_{t}+\textbf{v}\cdot\nabla) \mbox{\boldmath $\mu$} -\frac{\hbar}{2m\gamma}\partial^{\beta}[n\mbox{\boldmath $\mu$}, \partial^{\beta}\mbox{\boldmath $\mu$} ] +\mbox{\boldmath $\Im$}=\frac{2\gamma}{\hbar}n[\mbox{\boldmath $\mu$},\textbf{B}]. \end{equation}

This derivation shows that even in the self-consistent field approximation we have two unknown functions in the quantum hydrodynamic equations. These are the thermal pressure and the thermal part of the spin current. We call these parts the thermal parts since they are related to the distribution of particles on different quantum states. While the distribution in the form of Fermi step existing in fermions at the zero temperature also gives us some "thermal" parts. The Fermi pressure is one of well-known equations of state for the pressure. The Fermi pressure gives pressure of the unpolarized spin-1/2 fermions at zero temperature. In this paper we consider partially spin polarized electrons at temperatures mach smaller than their Fermi temperature. In our regime the pressure reads \cite{Andreev AoP 14 exchange}
\begin{equation}\label{SCES eq State single Fl polarised} \textrm{P}_{sf}=\frac{(3\pi^{2})^{\frac{2}{3}}}{10}\frac{\hbar^{2}}{m}\Biggl[\biggl(n+\frac{\textrm{M}_{z}}{\gamma}\biggr)^{\frac{5}{3}} +\biggl(n-\frac{\textrm{M}_{z}}{\gamma}\biggr)^{\frac{5}{3}}\Biggr], \end{equation}
where subindex "sf" reads single fluid. Here we refer to the single fluid model of electrons, while all plasma is considered as many fluid liquids and each species is considered as a fluid. Recently a two fluid model of spin-1/2 fermions was developed \cite{Andreev PRE 15 SEAW}, \cite{Andreev AoP 15 SEAW}, \cite{Andreev 1410 Comment}, where the spin-up and spin-down electrons are considered as two different fluids. We will apply this model to derivation of the thermal part of the spin current in the next section.

To the best of our knowledge it is impossible to give a straightforward derivation of the NLSE (\ref{SCES Non Lin Pauli Eq}) from the single fluid QHD, as it was done for the spinless regime \cite{MaksimovTMP 1999}. However we can justify the NLSE deriving the QHD equations from the NLSE as it was done in \cite{MaksimovTMP 2001}.

\section{\label{SCES sec:level1} Equation of state for the thermal part of the spin current}

In previous section we have demonstrated the existence of the thermal part of the spin current and formulated the problem of finding an equation of state for the thermal part of the spin current.

In terms of the macroscopic effective wave function $\Phi(\textbf{r},t)$ (\ref{SCES Non Lin Pauli Eq}) the particle concentration $n$, particle current $\textbf{j}$, and the magnetization $\textbf{M}$ appear as follows $n=\Phi^{+}\Phi$, $\textbf{j}=(\Phi^{+}\textbf{D}\Phi+(\textbf{D}\Phi^{+})\Phi)/2m$, $\textbf{M}=\gamma\Phi^{+}\widehat{\mbox{\boldmath $\sigma$}}\Phi$. Differentiating these functions with respect to time we find equations (\ref{SCES continuity equation with Velocity}), (\ref{SCES momentum balance eq with Velocity}), (\ref{SCES magnetric moment evol Eq with velocity}) derived from microscopic model in Sec. \ref{SCES sec:level1 Model}. Thus, we see that the NLSE (\ref{SCES Non Lin Pauli Eq}) is in agreement with the single fluid model of electrons. It is easy to see that the spinor pressure (\ref{SCES Pressure spinor}) introduced in NLSE (\ref{SCES Non Lin Pauli Eq}) leads to the agreement of equation (\ref{SCES Non Lin Pauli Eq}) with the two fluid model of electrons (the separate spin evolution QHD) developed in Refs. \cite{Andreev PRE 15 SEAW}, \cite{Andreev AoP 15 SEAW}, \cite{Andreev 1410 Comment}. However, in addition, we find an explicit form for the thermal part of the spin current
$$\Im^{\alpha}=\partial_{\beta}\textrm{J}^{\alpha\beta}_{th}=\gamma_{e}\frac{(6\pi^{2})^{2/3}\hbar}{m_{e}}(n_{\uparrow}^{2/3}-n_{\downarrow}^{2/3})\{S_{y},-S_{x}, 0\} $$
\begin{equation}\label{SCES spin current many part}  =\varepsilon^{\alpha\beta z}S^{\beta}\gamma_{e}\frac{(6\pi^{2})^{2/3}\hbar}{m_{e}}(n_{\uparrow}^{2/3}-n_{\downarrow}^{2/3}). \end{equation}
Let us rewrite formula (\ref{SCES spin current many part}) in the vector form and represent it in terms of the single fluid hydrodynamic model
$$\mbox{\boldmath $\Im$}=(\pi_{\uparrow}-\pi_{\downarrow})[\textbf{M}, \textbf{e}_{z}]/\hbar $$
\begin{equation}\label{SCES spin current many part Vector} =\frac{(3\pi^{2})^{2/3}\hbar }{m_{e}}\Biggl[\biggl(n-\frac{\textrm{M}_{z}}{\mid\gamma_{e}\mid}\biggr)^{2/3}-\biggl(n+\frac{\textrm{M}_{z}}{\mid\gamma_{e}\mid}\biggr)^{2/3}\Biggr] [\textbf{M}, \textbf{e}_{z}].\end{equation}

Formulae (\ref{SCES spin current many part}) and (\ref{SCES spin current many part Vector}) show that the term caused by the pressure in the NLSE (\ref{SCES Non Lin Pauli Eq}) reveals in the magnetic moment evolution equation as a torque of an effective force $\sim [\textbf{M}, \textbf{e}_{z}]$.

In the regime of the small magnetization limit formula (\ref{SCES spin current many part Vector}) transforms to
\begin{equation}\label{SCES} \mbox{\boldmath $\Im$}=-\frac{4(3\pi^{2})^{2/3}\hbar }{3m_{e}}\frac{\textrm{M}_{z}}{n^{1/3}}[\textbf{S}, \textbf{e}_{z}], \end{equation}
where we have used the spin density $\textbf{S}=\textbf{M}/\gamma$. In the small magnetization limit  the spin current $\mbox{\boldmath $\Im$}$ is proportional to the square of the spin density. As in the general case (\ref{SCES spin current many part Vector}), the spin current is proportional to $n^{5/3}$, since $\textbf{S}\sim n$.

The Landau-Lifshitz-Gilbert equation does not include the quantum part of the spin current \cite{Bottcher PRB 12}, \cite{Akhiezer PRE 98}. The convective part of the spin current is some times included in the Landau-Lifshitz-Gilbert equation: It is included, for instance, in Refs. \cite{Andreev PP 15 Positrons}, \cite{Akhiezer PRE 98}; as a recent example of the Landau-Lifshitz-Gilbert equation without the convective part of the spin current see \cite{Bottcher PRB 12}. Some generalizations of the Landau-Lifshitz-Gilbert equation are presented in literature. For instance, a nutation term is introduced in Ref. \cite{Bottcher PRB 12}, where it is substituted in the atomistic Landau-Lifshitz-Gilbert equation \cite{Skubic JP C}. The Landau-Lifshitz-Gilbert equation successfully describes the magnetic moment precession around and its relaxation towards the effective field acting on the magnetization on time scales down to femtoseconds \cite{Skomski JP C 03}, \cite{Hickey PRL 09}.

\subsection{Contribution of thermal part of spin current in quantum vorticity}

Considering quantum vorticity caused by the spin of particles, the quantum vorticity evolution equation has been found (see formula (13) in Ref. \cite{Mahajan PRL 11} and formula (4.6) in Ref. \cite{Takabayasi PTP 83}). The quantum vorticity has been combined with the classical vorticity to obtain the grand generalized vorticity \cite{Mahajan PRL 11}.

In this section we consider the contribution of the thermal part of spin current $\mbox{\boldmath $\Im$}$ in the grand generalized vorticity dynamics.

Quantum vorticity is defined in terms of the spin density $\textbf{S}$ and particle concentration $n$ \cite{Mahajan PRL 11}, \cite{Takabayasi PTP 83}
\begin{equation}\label{SCES vorticity definition} \Omega_{q}^{\alpha}=\frac{1}{2}\varepsilon^{\alpha\beta\gamma}\varepsilon^{\mu\nu\sigma} \biggl(\frac{S_{\mu}}{n}\biggr)\partial_{\beta}\biggl(\frac{S_{\nu}}{n}\biggr)\partial_{\gamma}\biggl(\frac{S_{\sigma}}{n}\biggr), \end{equation}
see for instance formula 12 and text below in Ref. \cite{Mahajan PRL 11} and formulae (4.10) and (4.11) in Ref. \cite{Takabayasi PTP 83}.

Differentiating definition (\ref{SCES vorticity definition}) with respect to time and applying the quantum hydrodynamic equations (\ref{SCES continuity equation with Velocity})-(\ref{SCES spin current many part Vector}) we find the quantum vorticity evolution equation
\begin{equation}\label{SCES} \partial_{t}\mbox{\boldmath $\Omega$}_{q}=\nabla\times(\textbf{v}\times\mbox{\boldmath $\Omega$}_{q})+\frac{q}{mc} \nabla\biggl(\frac{S_{\beta}}{n}\biggr) \times\nabla \hat{\textrm{B}}^{\beta}_{mod}, \end{equation}
where $\widehat{\textbf{B}}_{mod}=\widehat{\textbf{B}}+(\pi_{\downarrow}-\pi_{\uparrow})\textbf{e}_{z}/2\gamma$ and $\widehat{\textbf{B}}=\textbf{B}+(\hbar c/2qn)\partial^{\beta}(n\partial_{\beta}(\textbf{S}/n))$. We see that the effective magnetic field $\hat{\textbf{B}}$ is composed of the magnetic field $\textbf{B}$ and the contribution of the quantum Bohm potential in the magnetic moment evolution equation (the quantum part of the spin current). While the classical vorticity $\mbox{\boldmath $\Omega$}_{c}=\nabla\times(\textbf{A}+mc \textbf{v}/q)$ obeys the following equation
\begin{equation}\label{SCES} \partial_{t}\mbox{\boldmath $\Omega$}_{c}=\nabla\times(\textbf{v}\times\mbox{\boldmath $\Omega$}_{c})+\frac{q}{mc} \nabla\biggl(\frac{S_{\beta}}{n}\biggr) \times\nabla \hat{\textrm{B}}^{\beta}, \end{equation}
as it was shown in Refs. \cite{Mahajan PRL 11}, \cite{Takabayasi PTP 83}.

Constructing the grand generalized vorticities following Ref. \cite{Mahajan PRL 11} we have
\begin{equation}\label{SCES} \mbox{\boldmath $\Omega$}_{\pm}=\mbox{\boldmath $\Omega$}_{c}\pm\frac{\hbar c}{2q}\mbox{\boldmath $\Omega$}_{q} \end{equation}
and find that the thermal part of the spin current change the canonical form of vortex dynamics for $\mbox{\boldmath $\Omega$}_{-}$, since $\mbox{\boldmath $\Omega$}_{-}$ satisfy the following equation
\begin{equation}\label{SCES Omega - evol} \partial_{t}\mbox{\boldmath $\Omega$}_{-}=\nabla\times(\textbf{v}\times\mbox{\boldmath $\Omega$}_{-}) +\frac{c}{2q} \nabla\biggl(\frac{S_{z}}{n}\biggr)\times\nabla(\pi_{\uparrow}-\pi_{\downarrow}). \end{equation}
The first two terms in equation (\ref{SCES Omega - evol}) gives the canonical form of the vorticity evolution equation leading to the helicity conservation. The last term describes the contribution of the thermal part of the spin current. It is a generalization of equation obtained in Ref. \cite{Takabayasi PTP 83} (see formula (4.7)).

Takabayasi's works \cite{Takabayasi Soryusiron-Kenkyu 55}, \cite{Takabayasi PTP 222 55}, \cite{Takabayasi NC 56}, \cite{Takabayasi PR 56} provide us with the quantum vorticity research. The quantum relativistic vorticity (see formula (II) on page 17 of Ref. \cite{Takabayasi PTPS 57}) and corresponding four-vector Clebsch potential (the last relation in formula (f) on page 26 of Ref. \cite{Takabayasi PTPS 57}) can be found in Ref. \cite{Takabayasi PTPS 57}. 
Recent study of the vorticity structures can be found in Ref. \cite{Mahajan PP 15}. Quantum spirals explicitly arising from the Pauli equation are considered in Ref. \cite{Yoshida Mahajan arXiv 15}.



\section{\label{SCES sec:level1} Thermal part of the flux of the spin current}

The quantum hydrodynamic equations, as the classic hydrodynamics, are not restricted by the continuity, Euler and magnetic moment evolution equations. They can include the energy, pressure, thermal flux evolution equations, as it is well known from the classic five and thirteen moment models. In the spin-1/2 quantum plasmas we can derive the spin current evolution equation \cite{Andreev IJMP B 15}. This equation can be derived directly from the many-particle Schrodinger (Pauli) equation \cite{Andreev IJMP B 15}, or we can calculate moments of the distribution functions \cite{Zamanian PP 10}, \cite{Brodin PPCF 11}, \cite{Stefan PRE 11}, \cite{Andreev Phys A 15}. However the past derivations have not presented the thermal part of the flux of spin current, which existing in the spin current evolution equation.

In this section we present a derivation of the thermal part of the flux of spin current applying the NLSE (\ref{SCES Non Lin Pauli Eq}).

In terms of the many-particle effective wave function $\Phi(\textbf{r},t)$ (\ref{SCES Non Lin Pauli Eq}) the spin current arises as $\textrm{J}^{\alpha\beta}=\gamma(\Phi^{+}\textrm{D}^{\beta}\widehat{\sigma}^{\alpha}\Phi+(\textrm{D}^{\beta}\widehat{\sigma}^{\alpha}\Phi^{+})\Phi)/2m$. Applying this definition we derive the spin current evolution equation containing the thermal part of the spin current flux
$$\partial_{t}\textrm{J}^{\alpha\beta}+\partial^{\gamma}(\textrm{J}^{\alpha\beta}v^{\gamma})+\varepsilon^{\alpha\gamma z}\textrm{J}^{\gamma\beta}\frac{\pi_{\uparrow}-\pi_{\downarrow}}{\hbar}=\frac{q}{m}\textrm{M}^{\alpha}\textrm{E}^{\beta}$$
\begin{equation}\label{SCES spin current evolution} +\frac{q}{mc}\varepsilon^{\beta\gamma\delta}\textrm{J}^{\alpha\gamma}\textrm{B}^{\delta}+\frac{\gamma^{2}}{m}n\partial^{\beta}\textrm{B}^{\alpha} -\frac{2\gamma}{\hbar}\varepsilon^{\alpha\gamma\delta}\textrm{B}^{\gamma}\textrm{J}^{\delta\beta}\end{equation}
using the NLSE (\ref{SCES Non Lin Pauli Eq}) for the macroscopic wave function $\Phi$ time evolution.
The quantum part of the spin current flux was considered in Ref. \cite{Trukhanova PLA 15}.
The divergence of the thermal part of spin current flux is presented by the third term on the left-hand side. On the right-hand side of equation (\ref{SCES spin current evolution}), the contribution of the electromagnetic interaction in the spin current evolution is presented.

The obtained term is nonzero in the linear regime for the magnetized plasmas or magnetized dielectrics. Consequently, it gives a contribution in the wave dispersion.

If we want to derive an extended set of QHD equations containing the energy evolution \cite{MaksimovTMP 2001}, \cite{MaksimovTMP 1999}, pressure evolution, and spin current evolutions we do not have to use kinetic equations \cite{Zamanian PP 10}, \cite{Brodin PPCF 11}, \cite{Stefan PRE 11}, we can derive the hydrodynamic equations directly from the microscopic Schrodinger equation \cite{MaksimovTMP 2001}, \cite{Andreev IJMP B 15}, \cite{MaksimovTMP 1999} or some other methods \cite{Koide PRC 13}. Explicit contribution of the quantum Bohm potential in the energy evolution of spin-1/2 particles was derived in Ref. \cite{Trukhanova MPLB 15}.

\section{\label{SCES sec:level1} Spin waves in electron-ion plasmas}

Presence of the spin waves in plasmas and their dispersion was found in 2006 in Ref.
\cite{Vagin 06} by the application of the hybrid kinetic-hydrodynamic method, where the particle motion was described by the Vlasov equation, and the spin dynamics was described by the hydrodynamic like magnetic moment evolution equation. Waves propagating perpendicular to the external field were found there. The necessity of the anomalous magnetic moment of the electron was demonstrated there. Similar solution in terms of purely hydrodynamic description was found in Ref. \cite{Andreev VestnMSU 2007}. Since hydrodynamic description does not show the cyclotron resonances the anomalous magnetic moment was not include to distinguish the spin mode from the charge waves. Applying kinetics in the extended phase space, including two extra dimensions caused spin direction evolution similar spin wave was obtained in Ref. \cite{Brodin PRL 08}. Contribution of the quantum Bohm potential, existing in the magnetic moment evolution equation \cite{Takabayasi PTP 55}, in the spectrum of the spin waves was later found in Ref. \cite{Trukhanova PrETP 13}. Features of spin-plasma waves in the electron-positron plasmas arising due to the annihilation interaction in presence of the quantum Bohm potential were considered in Ref. \cite{Andreev PP 15 Positrons}.

Spin also gives contribution in longitudinal waves. For instance, it arises in the spectrum of Langmuir waves via the spin-orbit interaction \cite{Ivanov PTEP 15}.

We consider the high frequency oscillations, hence we assume that ions are motionless. We consider the electrons located in an external magnetic field $\textbf{B}_{0}=\textrm{B}_{0}\textbf{e}_{z}$. We are going to calculate the dispersion of the small amplitude wave excitations. The equilibrium state is the macroscopically motionless uniform systems of electrons, described by the equilibrium constant concentration $n_{0e}$, which is equal to the concentration of ions $n_{0i}$, zero velocity field $\textbf{v}_{0e}=0$, zero electric field $\textbf{E}_{0}=0$, and the equilibrium magnetization. If the magnetization is caused by the external magnetic field we can write $\textbf{M}_{0}=n_{0}\mbox{\boldmath $\mu$}_{0}=\chi \textbf{B}_{0}$, where $\chi$ is the ratio between equilibrium magnetic susceptibility and magnetic permeability. The small perturbations are described by $\delta n_{e}$, $\delta \textbf{v}_{e}$, $\delta \textbf{E}$, $\delta\mbox{\boldmath $\mu$}_{e}$ and $\delta \textbf{B}$.

After Fourier transformation of the linearized set of QHD equations (\ref{SCES continuity equation with Velocity})-(\ref{SCES spin current many part Vector}) we find the following set of algebraic equations
\begin{equation}\label{SCES} \omega\delta n-n_{0}\textbf{k}\delta \textbf{v}=0; \end{equation}
$$-\imath\omega\delta \textbf{v}+\imath \textbf{k} \tilde{v}_{Fe}^{2}\delta n +\imath \textbf{k} \frac{\hbar^{2}k^{2}}{4m^{2}}\delta n $$
\begin{equation}\label{SCES} =\frac{qn_{0}}{m}\delta \textbf{E}+\Omega_{e}n_{0}[\delta \textbf{v},\textbf{e}_{z}]+\frac{n_{0}\mu_{0}}{m}\imath \textbf{k}\delta \textrm{B}_{z}; \end{equation}
$$-\imath\omega\delta\mbox{\boldmath $\mu$}+\biggl(\frac{\hbar\mu_{0}}{2m\gamma}k^{2}+\textrm{w}\biggr)[\textbf{e}_{z}, \delta\mbox{\boldmath $\mu$}]$$
\begin{equation}\label{SCES}  =\frac{2\gamma}{\hbar}\textrm{B}_{0} [\delta\mbox{\boldmath $\mu$},\textbf{e}_{z}]+\frac{2\gamma}{\hbar}\mu_{0}[\textbf{e}_{z}, \delta \textbf{B}]; \end{equation}
and
$$k^{2}\delta \textbf{E}-\textbf{k}(\textbf{k}\textbf{E})=\frac{\omega^{2}}{c^{2}} \delta \textbf{E}$$
\begin{equation}\label{SCES}  +\frac{4\pi qn_{0}}{c^{2}}\imath\omega\delta \textbf{v}-\frac{4\pi\omega}{c}n_{0}[\textbf{k}, \delta\mbox{\boldmath $\mu$}]-\frac{4\pi\omega}{c}\mu_{0}\delta n [\textbf{k},\textbf{e}_{z}],\end{equation}
where
\begin{equation}\label{SCES}\tilde{v}_{Fe}^{2}= \frac{(3\pi^{2})^{\frac{2}{3}}\hbar^{2}}{6m^{2}}\biggl[\biggl(n_{0}+\frac{\textrm{M}_{0z}}{\gamma}\biggr)^{\frac{2}{3}} +\biggl(n_{0}-\frac{\textrm{M}_{0z}}{\gamma}\biggr)^{\frac{2}{3}}\biggr], \end{equation}
or in the small magnetization limit
\begin{equation}\label{SCES v Fe mod} \tilde{v}_{Fe}^{2}\approx \frac{1}{3}v_{Fe}^{2}\biggl[1-\frac{1}{9}\biggl(\frac{\textrm{M}_{z}}{\gamma n_{0}}\biggr)^{2}\biggr], \end{equation}
and
\begin{equation}\label{SCES w def}\textrm{w}=\frac{(3\pi^{2})^{\frac{2}{3}}\hbar}{m} \biggl[\biggl(n_{0}+\frac{\textrm{M}_{0z}}{\mid\gamma\mid}\biggr)^{\frac{2}{3}} -\biggl(n_{0}-\frac{\textrm{M}_{0z}}{\mid\gamma\mid}\biggr)^{\frac{2}{3}}\biggr]\end{equation}
arising from the equations of state for the degenerate electron gas (\ref{SCES eq State single Fl polarised}) and (\ref{SCES spin current many part Vector}). Frequency $\textrm{w}$ is the characteristic frequency for the thermal part of the spin current. Hence we call it the thermal spin current frequency (TSCF). It is necessary to compare the extra characteristic frequency $\textrm{w}$ with the Langmuir frequency, which is the major characteristic frequency in plasmas. Hence we present this comparison in Fig. \ref{SCES F TSCF vs Langm}. We note that TSCF arises in systems of neutral particles either.

We express all hydrodynamic variables via the amplitude of the electric field perturbation $\textbf{E}_{A}$ and obtain a set of three algebraic equations $\Lambda^{\alpha\beta}\textrm{E}_{A}^{\beta}=0$.

Our analysis is dedicated to point-like objects: electrons and positrons. Motionless ions are also considered as point-like objects. Model of the finite radius ions in electron-ion plasmas was considered in Ref. \cite{Andreev PP 15 FiniteSize}.


\subsection{Propagation parallel to the external field field}

Tensor $\Lambda^{\alpha\beta}$ can be separated on two parts
\begin{equation}\label{SCES disp matrix parall content}
\Lambda_{\alpha\beta}=\Lambda_{\alpha\beta}'+S_{\alpha\beta}.\end{equation}

The first of them related to the charge dynamics
\begin{equation}\label{SCES hat Lambda parallel}\hat{\Lambda}'=\left(\begin{array}{ccc}
\frac{\omega^{2}}{c^{2}}\Xi-k^{2}_{z}&
\frac{\omega^{2}_{e}}{c^{2}}\frac{-\imath\omega\Omega_{e}}{\omega^{2}-\Omega^{2}_{e}}&0\\
\frac{\omega^{2}_{e}}{c^{2}}\frac{\imath\omega\Omega_{e}}{\omega^{2}-\Omega^{2}_{e}}&
\frac{\omega^{2}}{c^{2}}\Xi-k^{2}_{z}&0\\
0&0&\frac{\omega^{2}}{c^{2}}(1-\frac{\omega^{2}_{Le}}{\omega^{2}-\frac{1}{3}v^{2}_{Fe}k^{2}_{z}})
\end{array}\right),\end{equation}
where $\Xi\equiv(1-\frac{\omega^{2}_{Le}}{\omega^{2}-\Omega^{2}_{e}})$, $\Omega_{e}=\frac{q_{e}B_{0}}{m_{e}c}$ is the cyclotron frequency for a charge in the magnetic field.

\begin{figure}
\includegraphics[width=8cm,angle=0]{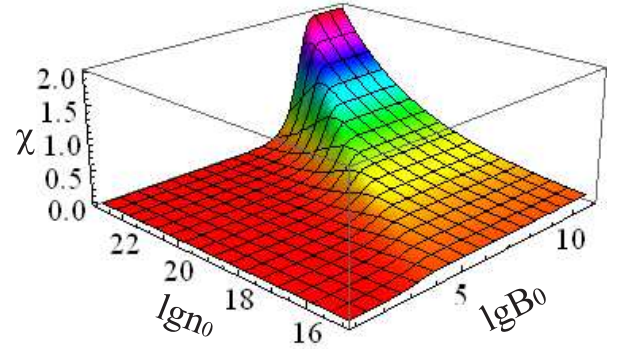}
\caption{\label{SCES F TSCF vs Langm} (Color online) This figure describes the ratio between the TSCF and the Langmuir frequency $\chi=\textrm{w}/\omega_{Le}$ as a function of the equilibrium concentration $\lg n_{0}$, where the concentration is measured in cm$^{-3}$, and the external magnetic field $\lg \textrm{B}_{0}$, where the magnetic field is measured in G.}
\end{figure}

\begin{figure}
\includegraphics[width=8cm,angle=0]{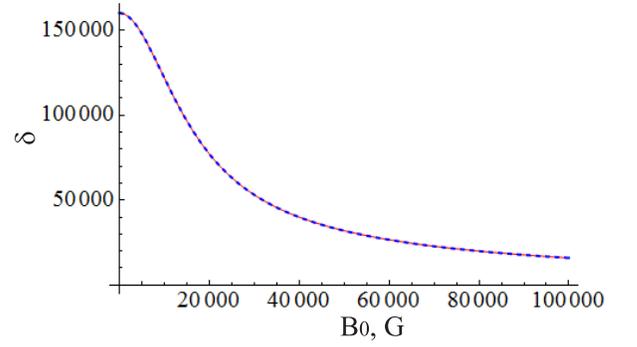}
\includegraphics[width=8cm,angle=0]{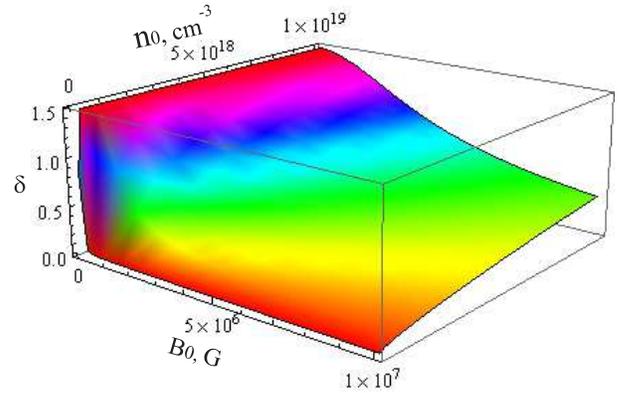}
\caption{\label{SCES F TSCF vs cyclotron} (Color online) This figure shows the ratio between the TSCF and the cyclotron frequency $\delta=\textrm{w}/\mid\Omega_{e}\mid$. The upper figure describes $\delta$ as a function of magnetic field at a fixed equilibrium concentration $n_{0}=10^{23}$ cm$^{-3}$. The lower figure presents $\delta$ as a function of the external magnetic field (large magnetic field) and the equilibrium concentration (small concentration).}
\end{figure}

The second part of tensor $\Lambda^{\alpha\beta}$ is caused by the spin dynamics
\begin{equation}\label{SCES hat S parallel}\hat{S}=\frac{k^{2}_{z}\omega_{\mu}}{\omega^{2}-\widetilde{\Omega}^{2}_{\gamma}} \left(\begin{array}{ccc} -\widetilde{\Omega}_{\gamma}&
-\imath\omega &0\\
\imath\omega &
-\widetilde{\Omega}_{\gamma} &0\\0&0&0
\end{array}\right),\end{equation}
where $\Omega_{\gamma}=\frac{2\gamma}{\hbar}B_{0}+\textrm{w}+\frac{\hbar^{2}k^{2}}{2m}\frac{\mu_{0}}{\gamma}$ is the generalized cyclotron frequency for a magnetic moment in the magnetic field, where the last term gives the quantum part of oscillation, the second term given by formula (\ref{SCES w def}) describes the collective effect caused by the Pauli principal, and $\omega_{\mu}=8\pi\gamma\mu_{0}n_{0}/\hbar$.

In formulae (\ref{SCES hat Lambda parallel}) and (\ref{SCES hat S parallel}) we have applied the modified Fermi velocity $\tilde{v}_{Fe}$ (\ref{SCES v Fe mod}) arising in accordance with the modified Fermi pressure (\ref{SCES eq State single Fl polarised}).

Nonzero perturbations exist if the determinant of matrix $\Lambda^{\alpha\beta}$ is equal to zero: $\det\hat{\Lambda}=0$. This condition gives the dispersion dependencies of wave perturbations.

General dispersion equation splits on two equations. One of them is for the longitudinal perturbations $\textbf{k}\parallel \textbf{e}_{z}\parallel\delta \textbf{E}$, and another one is for the transverse perturbations $\textbf{k}\perp\delta \textbf{E}$.

For the longitudinal excitations we find the spectrum of the Langmuir waves
\begin{equation}\label{SCES par pr Lengm wave}\omega^{2}=\omega^{2}_{Le}+\frac{1}{3}\tilde{v}^{2}_{Fe}k^{2}+\frac{\hbar^{2}k^{4}}{4m^{2}_{e}},\end{equation}
which contains the modified Fermi velocity. Which leads to the dependence of the Langmuir wave dispersion on the external magnetic field as it was mentioned in Ref. \cite{Andreev AoP 14 exchange} and numerically studied in Refs. \cite{Andreev PRE 15 SEAW}, \cite{Andreev AoP 15 SEAW}.

For the transverse waves we find the frequency dependence of the refractive index $n=kc/\omega$:
\begin{equation}\label{SCES par pr refractive index}\frac{k^{2}_{z}c^{2}}{\omega^{2}}=\frac{1-
\frac{\omega^{2}_{Le}}{\omega(\omega\mp\Omega_{e})}}{1\pm\frac{\omega_{\mu}}{\omega\mp\widetilde{\Omega}_{\gamma}}}.\end{equation}
Frequency $\omega_{\mu}$ can be rewritten as $\omega_{\mu}=4\pi\Omega_{e}\chi_{e}$, where $\chi_{e}$ is the ratio between equilibrium magnetic
susceptibility and magnetic permeability, as it was presented in Ref. \cite{Andreev VestnMSU 2007}.

Formula (\ref{SCES par pr refractive index}) can be presented as two equations for frequency
\begin{equation}\label{SCES tr waves parallel case 1} \omega^{2}-k_{z}^{2}c^{2}-\omega_{Le}^{2} \frac{\omega}{\omega-\Omega_{e}} -\frac{(k_{z}c)^{2}\omega_{\mu}}{\omega-\widetilde{\Omega}_{\gamma}}=0, \end{equation}
for the left circular polarized electric field $E_{y}=\imath E_{x}$, and
\begin{equation}\label{SCES tr waves parallel case 2} \omega^{2}-k_{z}^{2}c^{2}-\omega_{Le}^{2} \frac{\omega}{\omega+\Omega_{e}} +\frac{(k_{z}c)^{2}\omega_{\mu}}{\omega+\widetilde{\Omega}_{\gamma}}=0, \end{equation}
for the right circular polarized electric field $E_{y}=-\imath E_{x}$.

\begin{figure}
\includegraphics[width=8cm,angle=0]{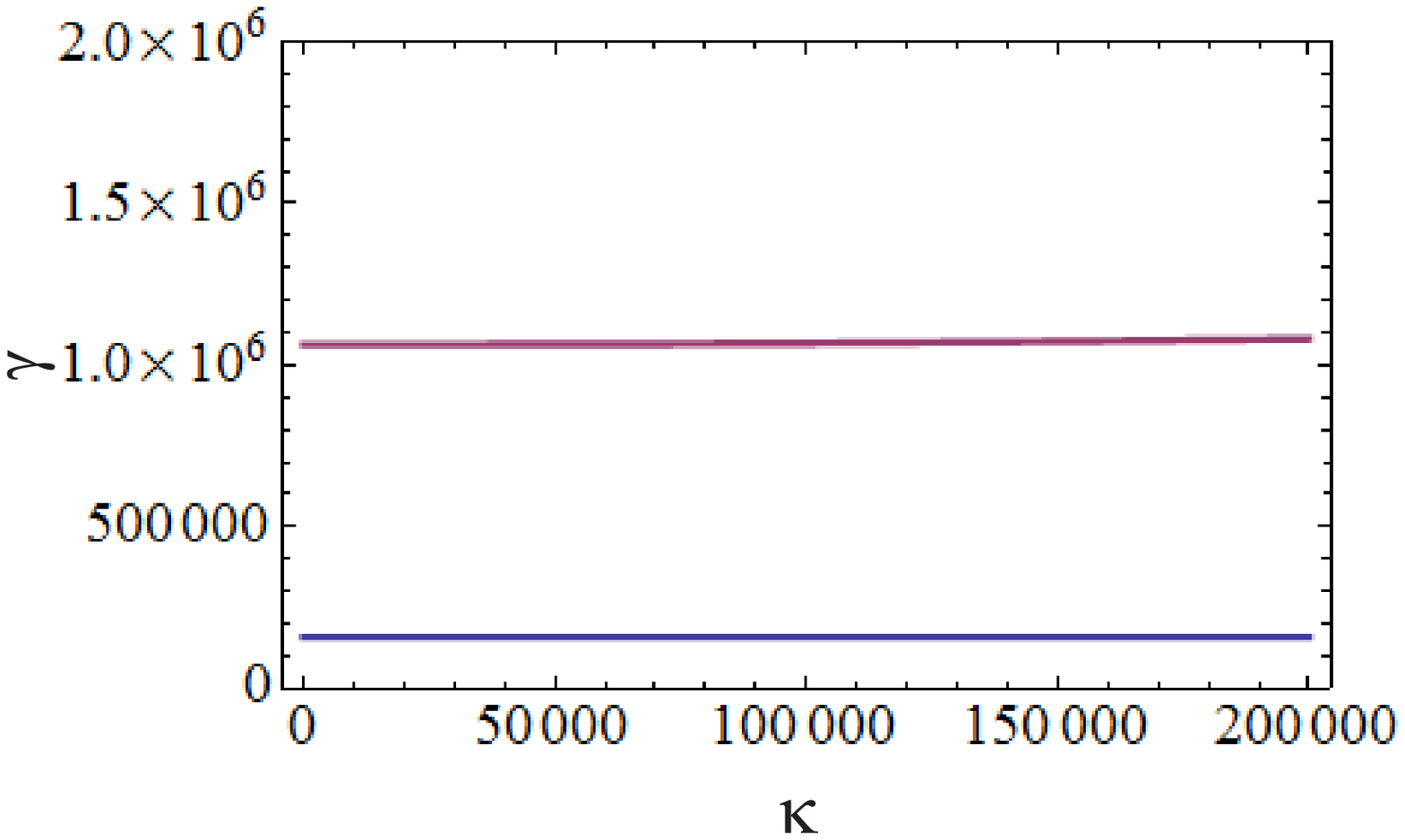}
\includegraphics[width=8cm,angle=0]{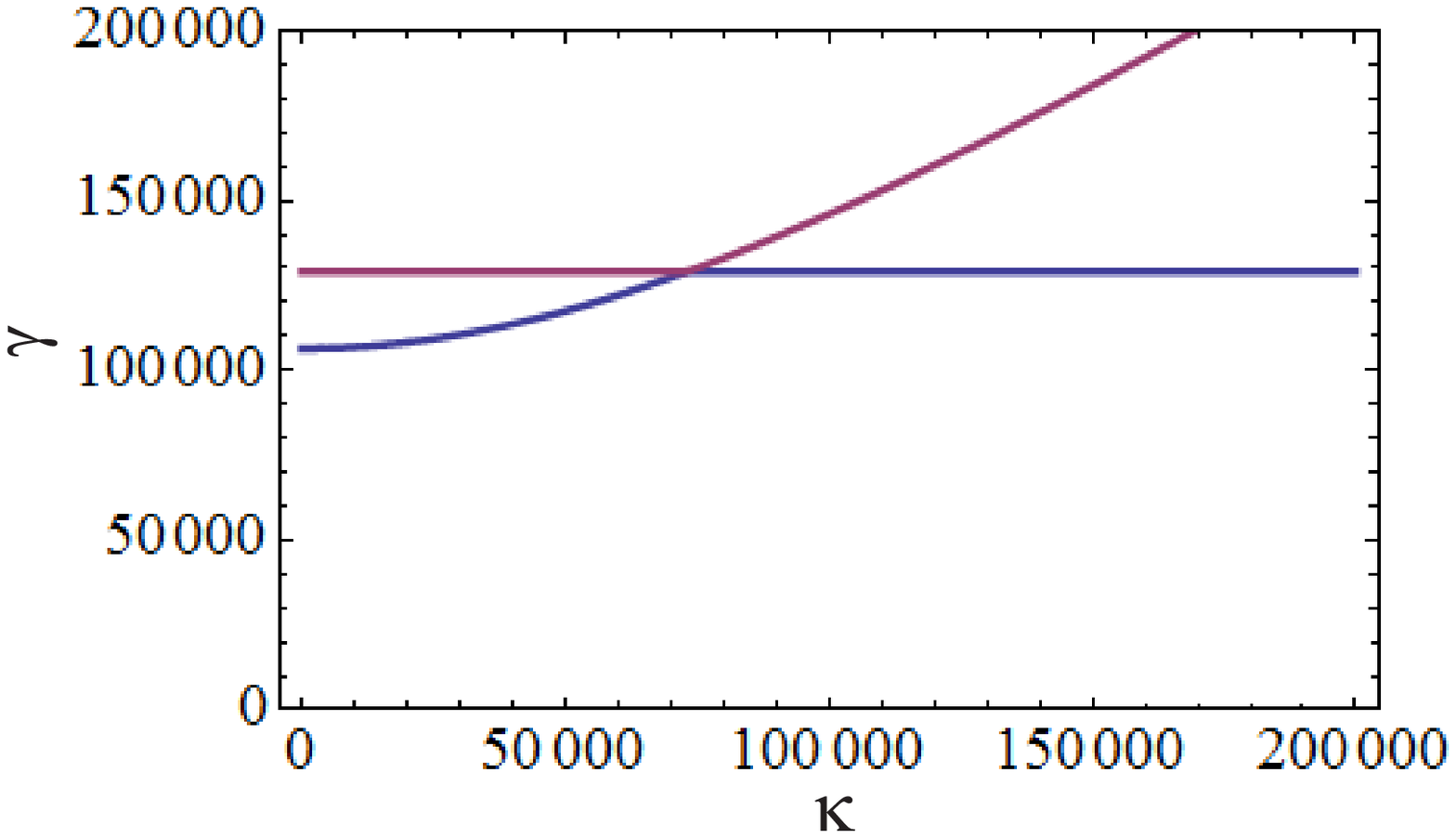}
\includegraphics[width=8cm,angle=0]{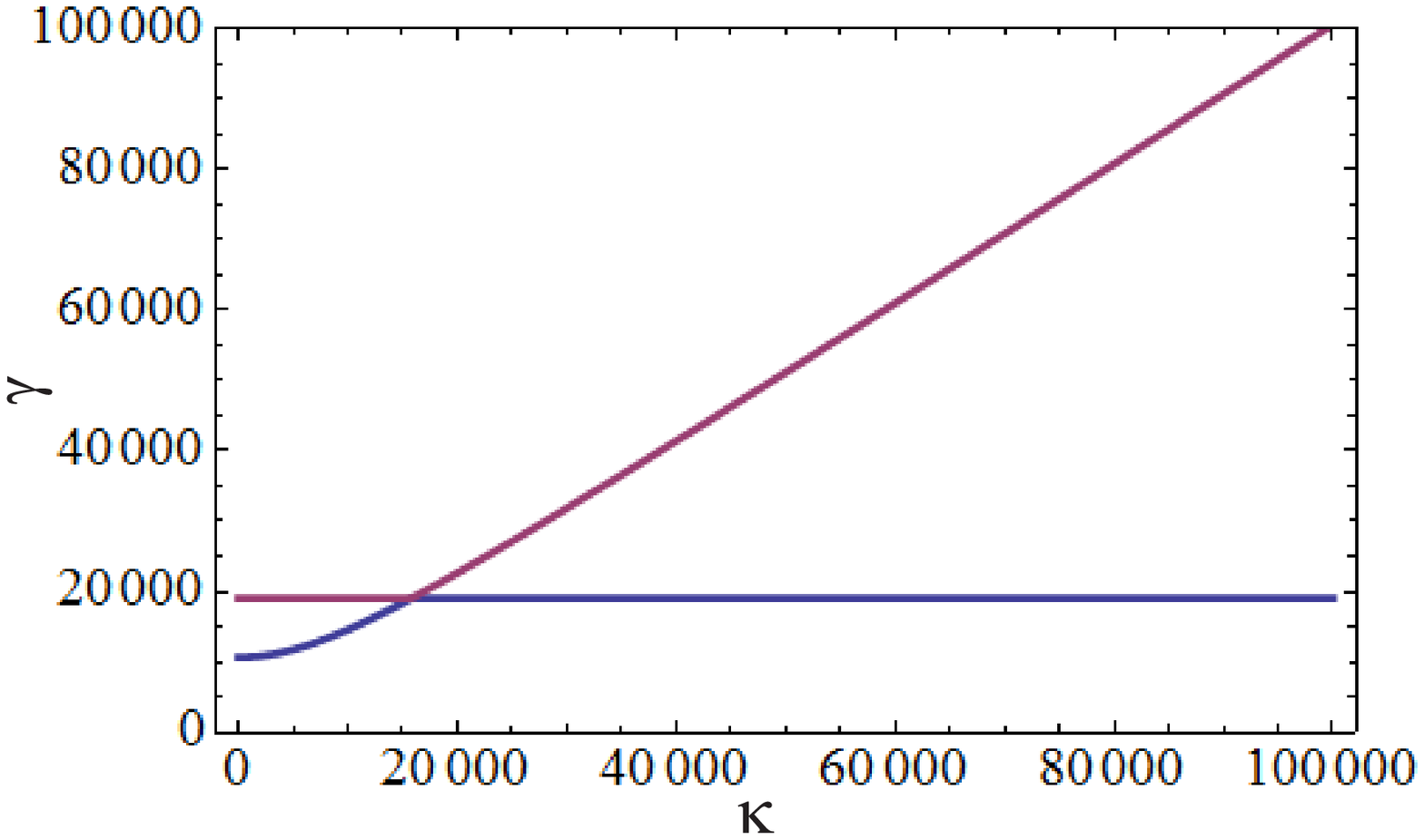}
\caption{\label{SCES F SpPlW and LW} (Color online) This figure presents solutions of equation (\ref{SCES tr waves parallel case 1}) giving frequency $\gamma=\omega/\mid\Omega_{e}\mid$ as  the functions of the dimensionless wave vector $\kappa=k_{z}c/\mid\Omega_{e}\mid=mc^{2}k_{z}/(eB_{0})$. Equation (\ref{SCES tr waves parallel case 1}) gives two solutions: the ordinary (electromagnetic) wave and the spin-plasma wave. At the equilibrium concentration $n_{0}=10^{23}$ cm$^{-3}$ and several values of the magnetic field we consider the frequency of these waves as a function of the wave vector ($\textrm{B}_{0}=10^{3}$ G, $\textrm{B}_{0}=10^{4}$ G, $\textrm{B}_{0}=10^{5}$ G). The upper (lower) line in the upper figure describes the ordinary (spin-plasma) wave. At the larger magnetic field these dispersion dependencies cross each other. Hence we have hybridization of these waves.}
\end{figure}

\begin{figure}
\includegraphics[width=8cm,angle=0]{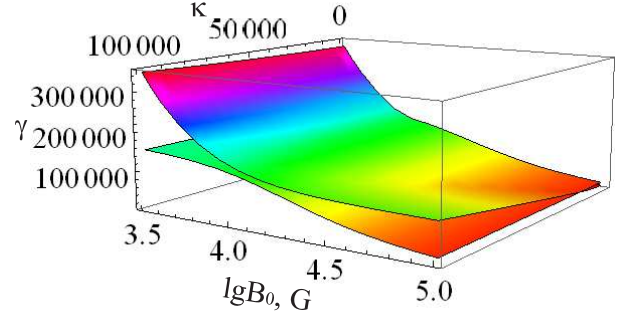}
\caption{\label{SCES F SpPlW and LW 3D} (Color online) This figure presents solutions of equation (\ref{SCES tr waves parallel case 1}) $\omega(k_{z})$ at the equilibrium concentration $n_{0}=10^{23}$ cm$^{-3}$ and continuously changing small magnetic field $\textrm{B}_{0}$. Here we use the dimensionless wave vector $\kappa=k_{z}c/\mid\Omega_{e}\mid=mc^{2}k_{z}/(eB_{0})$. It shows the hybridization of the ordinary electromagnetic and spin-plasma waves. }
\end{figure}

We start the analysis of the spin contribution in the circularly polarized electromagnetic/spin waves with the consideration of equation (\ref{SCES tr waves parallel case 1}).
Let us note that $\Omega_{e}=-\mid\Omega_{e}\mid$ since $q_{e}<0$, and $\widetilde{\Omega}_{\gamma}=-\mid\widetilde{\Omega}_{\gamma}\mid$ since $\gamma_{e}<0$, and we have $\mid\widetilde{\Omega}_{\gamma}\mid=2\mid\gamma_{e}\mid \textrm{B}_{0}/\hbar+(\mu_{0e}/\mid\gamma_{e}\mid)\hbar k^{2}/2m-\textrm{w}$ for small enough $\textrm{w}$. Equation (\ref{SCES tr waves parallel case 1}) can be rewritten as follows $\omega^{2}-k_{z}^{2}c^{2}-\omega\omega_{Le}^{2}/(\omega+\mid\Omega_{e}\mid)-\omega_{\mu}k_{z}^{2}c^{2}/(\omega+\mid\widetilde{\Omega}_{\gamma}\mid)=0$.
The last term changes the degree of this equation, which leads to extra solution. This solution is negative $\omega<0$, hence it does not have physical meaning. In the small wave vector regime, in the absence of the external magnetic field, but at the presence of inner effects creating the spin polarization of the conducting electrons, or at large TSCF, we have $\Omega_{\gamma}=\textrm{w}>0$, the comparison of the TSCF with the cyclotron frequency is presented in Fig. \ref{SCES F TSCF vs cyclotron}. This equation reappears as $\omega^{2}-k_{z}^{2}c^{2}-\omega\omega_{Le}^{2}/(\omega+\mid\Omega_{e}\mid)-\omega_{\mu}k_{z}^{2}c^{2}/(\omega-\textrm{w})=0$. In this regime, we find the spin plasma waves with the left circular polarization. Since $\omega_{\mu}\sim \textrm{M}_{0}$ is rather small, the last term gives noticeable at $\omega\simeq \textrm{w}$. We find solution as $\omega=\textrm{w}+\delta\omega$, where $\delta\omega\ll \textrm{w}$, so we have
\begin{equation}\label{SCES solution for TSCF} \omega=\textrm{w} +\frac{\omega_{\mu}k_{z}^{2}c^{2}}{\textrm{w}^{2}-k_{z}^{2}c^{2}-\frac{\textrm{w}\omega_{Le}^{2}}{\textrm{w}+\mid\Omega_{e}\mid}}. \end{equation}
As it follows from Fig. \ref{SCES F TSCF vs Langm}, for the large concentrations $n_{0}\sim 10^{23}$ cm$^{-3}$ and average magnetic fields $\textrm{B}_{0}<10^7$, the cyclotron frequency is negligible in compare with the TSCF. Moreover the TSCF is comparable with the Langmuir frequency, but the Langmuir frequency is larger than the TSCF $\omega_{Le}>\textrm{w}$. Hence, formula (\ref{SCES solution for TSCF}) can be simplified to $\omega=\textrm{w} -\omega_{\mu}k_{z}^{2}c^{2}/(k_{z}^{2}c^{2}+\omega_{Le}^{2}-\textrm{w}^{2})$.
If we do not make any assumptions about magnitude of the last term, which is caused by spin evolution, in the dispersion equation we need to solve it numerically. Solutions are presented in Figs. \ref{SCES F SpPlW and LW}, \ref{SCES F SpPlW and LW 3D}.

\begin{figure}
\includegraphics[width=8cm,angle=0]{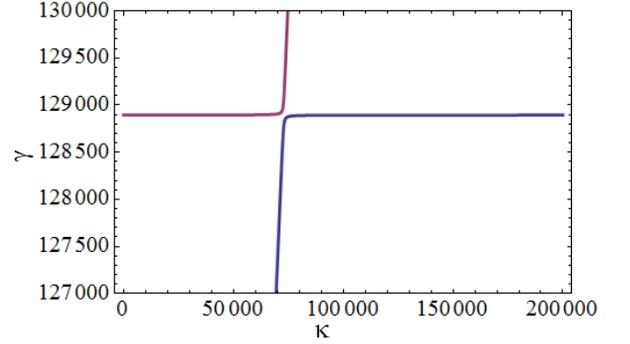}
\caption{\label{SCES F SpPlW and LW small Scale} (Color online) This figure shows in more details the hybridization of the ordinary  and spin-plasma waves at $\textrm{B}_{0}=10^{4}$ G and $n_{0}=10^{23}$ cm$^{-3}$ presented in middle picture in Fig. \ref{SCES F SpPlW and LW}.}
\end{figure}

At small wave vectors $k$ the frequency of the ordinary wave is comparable with the Langmuir frequency. If the TSCF is comparable we find the hybridization of the ordinary wave and the spin-plasma wave depicted in Figs. \ref{SCES F SpPlW and LW}, \ref{SCES F SpPlW and LW 3D}. In middle and lower pictures in Fig. \ref{SCES F SpPlW and LW} we see the crossing of the dispersion dependencies of the ordinary wave and the spin-plasma wave. To sight the hybridization we need to decrease the scale of frequencies as it is shown in Fig. \ref{SCES F SpPlW and LW small Scale} at $\textrm{B}_{0}=10^{4}$ G.

Let us consider equation (\ref{SCES tr waves parallel case 2}) describing the transverse waves with the right circularly polarization. Due to the negative charge of electrons we have $\omega^{2}-k_{z}^{2}c^{2}-\omega\omega_{Le}^{2}/(\omega-\mid\Omega_{e}\mid)-\omega_{\mu}k_{z}^{2}c^{2}/(\omega-\mid\widetilde{\Omega}_{\gamma}\mid)=0$ for small $\textrm{w}$, so $\mid\widetilde{\Omega}_{\gamma}\mid=2\mid\gamma_{e}\mid \textrm{B}_{0}/\hbar+(\mu_{0e}/\mid\gamma_{e}\mid)\hbar k^{2}/2m-\textrm{w}$. Since $\mid\widetilde{\Omega}_{\gamma}\mid\neq\mid\Omega_{e}\mid$ we see that the last term gives an extra positive solution, in compare with the spinless case, for the right circularly polarized waves. This solution has frequency near $\mid\widetilde{\Omega}_{\gamma}\mid$. Hence, we obtain
\begin{equation}\label{SCES RCP SPW} \omega=\mid\widetilde{\Omega}_{\gamma}\mid -\frac{\omega_{\mu}k_{z}^{2}c^{2}}{\widetilde{\Omega}_{\gamma}^{2}-k_{z}^{2}c^{2}-\omega_{Le}^{2} \frac{\mid\widetilde{\Omega}_{\gamma}\mid}{\mid\widetilde{\Omega}_{\gamma}\mid-\mid\Omega_{e}\mid}}\end{equation}
where difference $\mid\widetilde{\Omega}_{\gamma}\mid-\mid\Omega_{e}\mid$ is equal to $(g-1)e\textrm{B}_{0}/mc+\mu_{0z}k_{z}^{2}c/ge-\textrm{w}$. If contribution of $\textrm{w}$ dominates in $\widetilde{\Omega}_{\gamma}$ equation (\ref{SCES tr waves parallel case 2}) does not have extra solution.

To conclude this subsection we note that if we do not account the thermal part of the spin current or assume it to be rather small we find the spin-plasma wave as a part of the right circularly polarized wave spectrum. If the thermal part of the spin current dominates over the cyclotron frequency and the quantum part of the spin current the right circularly polarized spin-plasma wave does not exist, but we find a left circularly polarized spin-plasma wave. For the large enough TPSC we find hybridization of the spin-plasma wave and the ordinary electromagnetic wave.

\subsection{Propagation perpendicular to the external field field}

Having the generalization of spin-1/2 QHD developed in 2001 \cite{MaksimovTMP 2001} we are going to generalize the spin plasma wave spectrum calculated in Refs. \cite{Andreev VestnMSU 2007}, \cite{Vagin 06}.

In the regime of perpendicular propagation we have $k_{z}=0$. Aiming to study the transverse waves we work with element $\Lambda^{zz}$ of the dispersion matrix $\widehat{\Lambda}$. Thus, our dispersion equation is
\begin{equation}\label{SCES Lambda zz tr} \Lambda^{zz}=\frac{\omega^{2}}{c^{2}}-k_{\perp}^{2}-\frac{\omega^{2}_{Le}}{c^{2}} -\frac{\omega_{\mu}\widetilde{\Omega}_{\gamma}k_{\perp}^{2}}{\omega^{2}-\widetilde{\Omega}^{2}_{\gamma}}=0.\end{equation}
Using representation $\omega_{\mu}=4\pi\Omega_{e}\chi_{e}$, as it was done in Ref. \cite{Andreev VestnMSU 2007}, the numerator can be rewritten as $4\pi\Omega^{2}_{e}k_{\perp}^{2}\chi_{e}$.

Dispersion equation (\ref{SCES Lambda zz tr}) gives two solutions. Their general form can be presented analytically:
$$\omega^{2}=\frac{1}{2}\biggl[\omega^{2}_{Le} +k_{\perp}^{2}c^{2} +\widetilde{\Omega}^{2}_{\gamma}$$
\begin{equation}\label{SCES Lambda zz tr general solution}  \pm\sqrt{(\omega^{2}_{Le} +k_{\perp}^{2}c^{2} -\widetilde{\Omega}^{2}_{\gamma})^{2}+4\omega_{\mu}\widetilde{\Omega}_{\gamma}k_{\perp}^{2}c^{2}}\biggr].\end{equation}

Solving equation (\ref{SCES Lambda zz tr}) for the frequencies far from the shifted cyclotron frequency $\mid\widetilde{\Omega}_{\gamma}\mid$, and applying the iteration method, we find the contribution of the spin dynamics in the dispersion of the electromagnetic (ordinary) waves
\begin{equation}\label{SCES Lambda zz tr old rel}\omega^{2}=\omega^{2}_{Le}+k_{\perp}^{2}c^{2} +\frac{k_{\perp}^{2}c^{2}\omega_{\mu}\widetilde{\Omega}_{\gamma}}{\omega^{2}_{Le}+k_{\perp}^{2}c^{2}-\widetilde{\Omega}^{2}_{\gamma}}.\end{equation}

Considering frequencies near $\mid\widetilde{\Omega}_{\gamma}\mid$ we obtain
\begin{equation}\label{SCES Lambda zz tr new rel}\omega=|\widetilde{\Omega}_{\gamma}|\biggl(1 -\frac{\omega_{\mu}}{2\widetilde{\Omega}_{\gamma}}\frac{
k_{\perp}^{2}c^{2}}{\omega^{2}_{Le}+k_{\perp}^{2}c^{2}-\widetilde{\Omega}^{2}_{\gamma}}\biggr).\end{equation}

Solutions (\ref{SCES Lambda zz tr old rel}) and (\ref{SCES Lambda zz tr new rel}) generalize solutions (8) and (9) found in Ref. \cite{Andreev VestnMSU 2007}.

The TSCF can dominate in $\widetilde{\Omega}_{\gamma}$ in the regime of the perpendicular propagation. It reveals in the spectrum similar to the spectrum found and described for the parallel propagation.

The fluid moment hierarchy was addressed in Refs. \cite{Zamanian PP 10}, \cite{Brodin PPCF 11}, \cite{Stefan PRE 11} at the application of the generalized Wigner kinetics to the spin evolution in quantum plasmas. The spinless case was considered in Refs. \cite{Haas NJP 10}, \cite{Haas PLA 09}. The account of higher moments in hydrodynamics of spin-1/2 particles generalizes the dispersion equation for the transverse waves and gives more resonance terms (see equation (16) in Ref. \cite{Zamanian PP 10}). That leads to resonances close to double the cyclotron frequency and close to the difference of the cyclotron frequency of charge and the cyclotron frequency of magnetic moment in the external magnetic field, the last one is shifted by the anomalous magnetic moment.

The TPSC modifies the cyclotron frequency arising from the magnetic moment precession. Therefore, effects described in the previous paragraph can be also affected by the TPSC.

\section{\label{SCES sec:level1} Features of QHD model for spin-1/2 electron-positron plasmas}

The model of spin-1/2 electron-positron plasmas differs from the model of spin-1/2 electron ion plasmas due to the existence of the additional interaction between electrons and positrons called the annihilation interaction. Corresponding hydrodynamic and kinetic models were recently developed in Ref. \cite{Andreev PP 15 Positrons}.

In this paper we generalize the hydrodynamic model developed in Ref. \cite{Andreev PP 15 Positrons} including the thermal part of the spin current.

Hamiltonian of the electron-positron plasmas differs from Hamiltonian (\ref{SCES Hamiltonian gen}), as it was demonstrated in Ref. \cite{Andreev PP 15 Positrons}. The difference arises from the existence of the annihilation interaction. The spinless part of the annihilation interaction is similar to the Darwin interaction. Therefore, we include these interactions together:
$$\Delta \hat{\textrm{H}}=-\frac{1}{2}\sum_{i,j\neq i}^{N}\frac{\pi q_{i}q_{j}\hbar^{2}}{m^{2}c^{2}}\delta(\textbf{r}_{i}-\textbf{r}_{j})$$
\begin{equation}\label{SCES Hamiltonian with positrons}-\sum_{i=1}^{N_{e-}}\sum_{j=N_{e-}+1}^{N_{e-}+N_{e+}}\frac{\pi q_{i}q_{j}\hbar^{2}}{2m^{2}c^{2}}(3+\mbox{\boldmath $\sigma$}_{i}\cdot\mbox{\boldmath $\sigma$}_{j})\delta(\textbf{r}_{i}-\textbf{r}_{j}),\end{equation}
where the first term is the Darwin interaction between all particles (the electron-electron, positron-positron, and electron-positron interactions), and the second term is the annihilation interaction between electrons and positrons.

Choosing the coefficient in the Hamiltonian of the Darwin interaction we follow the Breit Hamiltonian for two electrons or the pair of electron and positron following from the quantum electrodynamic scattering amplitude instead of the Darwin term arising from the Dirac equation for the single electron in the external field (see book \cite{Landau 4} and discussion in Refs. \cite{Ivanov PTEP 15}, \cite{Andreev RPJ 14 Rel2D}).

The relativistic part of the kinetic energy gives a contribution similar to the Darwin interaction. Hence it should be included at the relativistic analysis of some effects \cite{Andreev PP 15 Positrons}, \cite{Ivanov PTEP 15}. However, the spectrum of spin-plasma waves is not affected by this effect, so we do not consider it here.

In the accordance with the method of many-particle quantum hydrodynamics described above we find the continuity equations for the electrons
\begin{equation}\label{SCES continuity eq for electron} \partial_{t}n_{e}+\nabla\cdot(n_{e}\textbf{v}_{e})=0, \end{equation}
and the positrons
\begin{equation}\label{SCES continuity eq for positron} \partial_{t}n_{p}+\nabla\cdot(n_{p}\textbf{v}_{p})=0, \end{equation}
the Euler equations for the electrons
$$mn_{e}(\partial_{t}+\textbf{v}_{e}\cdot\nabla)\textbf{v}_{e}+\nabla \textrm{P}_{e}-\frac{\hbar^{2}}{4m}n_{e}\nabla\Biggl(\frac{\triangle n_{e}}{n_{e}}-\frac{(\nabla n_{e})^{2}}{2n_{e}^{2}}\Biggr)$$
$$+\frac{\hbar^{2}}{4m\gamma^{2}_{e}}\partial^{\beta}\biggl(n_{e}(\partial^{\beta}\mu_{e}^{\gamma})\nabla\mu_{e}^{\gamma}\biggr)
=q_{e}n_{e}\textbf{E}+\frac{q_{e}}{c}n_{e}[\textbf{v}_{e},\textbf{B}]$$
\begin{equation}\label{SCES Euler equation for electrons}
+n_{e}\mu^{\beta}_{e}\nabla (\textrm{B}^{\beta} +2\pi n_{p}\mu^{\beta}_{p})+\frac{\pi q_{e}\hbar^{2}}{m^{2}c^{2}}n_{e} \nabla (q_{e}n_{e}+\frac{5}{2}q_{p}n_{p}),\end{equation}
and for the positrons
$$mn_{p}(\partial_{t}+\textbf{v}_{p}\cdot\nabla)\textbf{v}_{p}+\nabla \textrm{P}_{p}-\frac{\hbar^{2}}{4m}n_{p}\nabla\Biggl(\frac{\triangle n_{p}}{n_{p}}-\frac{(\nabla n_{p})^{2}}{2n_{p}^{2}}\Biggr)$$
$$+\frac{\hbar^{2}}{4m\gamma^{2}_{p}}\partial^{\beta}\biggl(n_{p}(\partial^{\beta}\mu_{p}^{\gamma})\nabla\mu_{p}^{\gamma}\biggr)
=q_{p}n_{p}\textbf{E}+\frac{q_{p}}{c}n_{p}[\textbf{v}_{p},\textbf{B}]$$
\begin{equation}\label{SCES Euler equation for positrons}+n_{p}\mu^{\beta}_{p}\nabla (\textrm{B}^{\beta} +2\pi n_{e}\mu^{\beta}_{e})+\frac{\pi q_{p}\hbar^{2}}{m^{2}c^{2}}n_{p} \nabla (q_{p}n_{p}+\frac{5}{2}q_{e}n_{e}),\end{equation}
where we have used the reduced magnetization $\mbox{\boldmath $\mu$}_{a}$ defined by the following formula
$\textbf{M}_{a}(\textbf{r},t)=n_{a}\mbox{\boldmath $\mu$}_{a}$.

Most of the terms in Euler equations (\ref{SCES Euler equation for electrons}) and (\ref{SCES Euler equation for positrons}) have same meaning as similar terms in the Euler equation (\ref{SCES momentum balance eq with Velocity}) described after equation (\ref{SCES magnetric moment evol Eq with velocity}). However, equations (\ref{SCES Euler equation for electrons}) and (\ref{SCES Euler equation for positrons}) contain some new terms specific for electron-positron interaction. Let us describe them here. These terms follows from additional part to the Hamiltonian (\ref{SCES Hamiltonian with positrons}).
We have four groups of terms on the right-hand side of each Euler equation. The account of Hamiltonian (\ref{SCES Hamiltonian with positrons}) leads to the second part of the third group of terms and to the existence of the fourth group. The contribution to the third group  caused by the spin dependent part of the annihilation interaction. The first part of the fourth term is the Darwin interaction of the particles of the same species. Its second part is the combination of the interspecies Darwin interaction (1 of 5/2) and the spinless part of the annihilation interaction (3/2 of 5/2). The pressures $\textrm{P}_{e}$ and $\textrm{P}_{p}$ are given by formula (\ref{SCES eq State single Fl polarised}).

We also find the magnetic moment evolution equations for the electrons
$$n_{e}(\partial_{t}+\textbf{v}_{e}\cdot\nabla) \mbox{\boldmath $\mu$}_{e} -\frac{\hbar}{2m\gamma_{e}}\partial^{\beta}[n_{e} \mbox{\boldmath $\mu$}_{e}, \partial^{\beta}\mbox{\boldmath $\mu$}_{e}] +\mbox{\boldmath $\Im$}_{e}$$
\begin{equation}\label{SCES eq of magnetic moments evol electr} =\frac{2\gamma_{e}}{\hbar}n_{e}[\mbox{\boldmath $\mu$}_{e},\textbf{B}]+ \frac{4\pi\gamma_{e}}{\hbar} n_{e}n_{p}[\mbox{\boldmath $\mu$}_{e}, \mbox{\boldmath $\mu$}_{p}],\end{equation}
and for the positrons
$$n_{p}(\partial_{t}+\textbf{v}_{p}\cdot\nabla) \mbox{\boldmath $\mu$}_{p}-\frac{\hbar}{2m\gamma_{p}}\partial^{\beta}[n_{p}
\mbox{\boldmath $\mu$}_{p},\partial^{\beta}\mbox{\boldmath $\mu$}_{p}] +\mbox{\boldmath $\Im$}_{p}$$
\begin{equation}\label{SCES eq of magnetic moments evol positr} =\frac{2\gamma_{p}}{\hbar}n_{p}[\mbox{\boldmath $\mu$}_{p},\textbf{B}]+ \frac{4\pi\gamma_{p}}{\hbar} n_{e}n_{p}[\mbox{\boldmath $\mu$}_{p}, \mbox{\boldmath $\mu$}_{e}].\end{equation}
Equations (\ref{SCES eq of magnetic moments evol electr}) and (\ref{SCES eq of magnetic moments evol positr}) differ from equation (\ref{SCES magnetric moment evol Eq with velocity}) by the last term, which is caused by the spin part of the annihilation interaction (see Hamiltonian (\ref{SCES Hamiltonian with positrons})). Vectors $\mbox{\boldmath $\Im$}_{e}$ and $\mbox{\boldmath $\Im$}_{p}$ are given by formula (\ref{SCES spin current many part Vector}).

\section{Spin waves in electron-positron plasmas}

Spin-plasma waves in electron-positron plasmas show some similarity to the electron-ion spin-plasma waves. Principal difference is in the existence of the annihilation interaction.

\subsection{Propagation parallel to the external field field}

Looking on matrices (\ref{SCES hat Lambda parallel}) and (\ref{SCES hat S parallel}) we see that at consideration of the electron-positron plasmas the non-diagonal elements disappear since the contributions of electrons and positrons cancel each other. Therefore the circularly polarized, in the electron-ion plasmas, wave are the linearly polarized in the electron-positron plasmas and the account of the spin dynamics does not change it. Nevertheless, the spin gives a contribution in the dispersion of the electromagnetic linearly polarized waves. Moreover, the spin leads to the spin-plasma waves similarly to the electron-ion plasmas described above.

The dispersion equation for the transverse waves propagating perpendicular to the external magnetic field with the account of the spin evolution together with the contribution of the thermal part of the spin current arises as follows
$$k_{z}^{2}c^{2}-\omega^{2}+2\omega_{Le}^{2} \frac{\omega^{2}}{\omega^{2}-\Omega^{2}_{e}}$$
\begin{equation}\label{SCES ep transverse wave with annihil} +\frac{2\mid\gamma\mid}{\hbar}\frac{8\pi(k_{z}c)^{2}n_{0}\mu_{0}(\Theta+\Lambda)}{\omega^{2}-\Theta^{2}+\Lambda^{2}}=0, \end{equation}
where
\begin{equation}\label{SCES Theta def ep} \Theta=\frac{2\gamma}{\hbar}B_{0}+\frac{\hbar\mu_{0}}{2m\gamma}k_{z}^{2}+\frac{4\pi\gamma}{\hbar} n_{0}\mu_{0}+\textrm{w},\end{equation}
and
\begin{equation}\label{SCES Lambda def ep} \Lambda=4\pi\mid\gamma\mid n_{0}\mu_{0}/\hbar.\end{equation}
Frequency $\Theta$, given by formula (\ref{SCES Theta def ep}), is the shifted cyclotron frequency for a magnetic moment in the external magnetic field, it is shifted, from the traditional cyclotron frequency $e\textrm{B}_{0}/mc$, due to the anomalous magnetic moment of the electrons and positrons included in $\gamma$, the quantum Bohm potential contribution in the magnetic moment evolution equation (the second term in (\ref{SCES Theta def ep})), the spin dependent part of the annihilation interaction (the third term), and, found in this paper, contribution of the thermal part of the spin current $\textrm{w}$ given by formula (\ref{SCES w def}). Frequency (\ref{SCES Lambda def ep}) is the characteristic frequency of the spin dependent part of the annihilation interaction.

The contribution of the thermal part of the spin current in the resonance frequency for electron-ion plasmas is described above. The electron-positron plasmas differs by the presence of the annihilation interaction, which increases frequency $\Theta$ on a constant $\Lambda$ (\ref{SCES Lambda def ep}).

The frequency of the spin-plasma wave is close to the resonance frequency $\omega_{R}=\sqrt{\Theta^{2}-\Lambda^{2}}$. Let us consider this frequency in more details. The annihilation interaction arises in the resonance frequency $\sqrt{\Theta^{2}-\Lambda^{2}}$ twice. These contributions partially cancel each other. Considering square of $\Theta$ explicitly we have $\sqrt{\Theta^{2}-\Lambda^{2}}= \sqrt{(\frac{2\gamma}{\hbar}B_{0}+\frac{\hbar\mu_{0}}{2m\gamma}k_{z}^{2}+\textrm{w}) (\frac{2\gamma}{\hbar}B_{0}+\frac{\hbar\mu_{0}}{2m\gamma}k_{z}^{2}+\textrm{w}+2\Lambda)}$. It shows that the annihilation interaction increases the resonance frequency, while the thermal part of the spin current decreases it, as it has  been demonstrated in more details for the electron-ion plasmas.

\subsection{Propagation perpendicular to the external field field}

In this regime the dispersion equation for the transverse waves arises as follows
\begin{equation}\label{SCES ep Tr DE Ez}\omega^{2}-k_{x}^{2}c^{2}-2\omega_{Le}^{2} -8\pi n_{0}\mu_{0}\frac{2\gamma}{\hbar}\frac{k_{x}^{2}c^{2}(\Theta+\Lambda)}{\omega^{2}-\Theta^{2}+\Lambda^{2}}=0. \end{equation}
The last term in equation (\ref{SCES ep Tr DE Ez}) arises due to the spin of the electrons and positrons. It increases the degree of the dispersion equation and leads to the spin-plasma wave appearance, as it was also demonstrated in Sect VI B for the electron-ion plasmas.

If contribution of the magnetization in equation (\ref{SCES ep Tr DE Ez}) is small the spin-plasma wave arises at frequencies close to the resonance frequency of the last term in formula (\ref{SCES ep Tr DE Ez}) $\omega\approx\sqrt{\Theta^{2}-\Lambda^{2}}$. Explicitly including the small shift from the resonance frequency we find the dispersion of the spin-plasma wave propagating perpendicular to the external field:
\begin{equation}\label{SCES ep spin wave perpendicular}  \omega=\sqrt{\Theta^{2}-\Lambda^{2}}\biggl(1-\frac{1}{\Theta-\Lambda}\frac{8\pi\mid\gamma\mid n_{0}\mu_{0}k_{x}^{2}c^{2}/\hbar}{2\omega_{Le}^{2}+k_{x}^{2}c^{2}+\Lambda^{2}-\Theta^{2}}\biggr).\end{equation}
It is an analog of solution (\ref{SCES Lambda zz tr new rel}) found for the electron-ion plasmas.

At the parallel and perpendicular propagation of the spin-plasma waves we have the same resonance frequency $\sqrt{\Theta^{2}-\Lambda^{2}}$.

Formula (\ref{SCES ep spin wave perpendicular}) gives an approximate solution of equation (\ref{SCES ep Tr DE Ez}) for the spin-plasma wave. Besides, we can present general analytical solution of equation (\ref{SCES ep Tr DE Ez}) describing the electromagnetic and spin-plasma waves together:
$$\omega^{2}=\frac{1}{2}\biggl[2\omega^{2}_{Le} +k_{\perp}^{2}c^{2} +\Theta^{2}-\Lambda^{2}$$
\begin{equation}\label{SCES ep Lambda zz tr general solution}  \pm\sqrt{(2\omega^{2}_{Le} +k_{\perp}^{2}c^{2} -\Theta^{2}+\Lambda^{2})^{2}+8\omega_{\mu}(\Theta+\Lambda)k_{\perp}^{2}c^{2}}\biggr].\end{equation}
This solution includes the hybridization of two branches studied numerically for waves propagating parallel to the external field in the electron-ion plasmas and presented in Figs. \ref{SCES F SpPlW and LW 3D} and \ref{SCES F SpPlW and LW small Scale}.

\section{\label{SCES sec:level1} Conclusion}

The thermal part of the spin current has not been considered in the condensed matter physics and physics of quantum plasmas. In this paper we have improved the magnetic moment evolution equation for degenerate electrons and applied it for the spin-1/2 quantum plasma phenomena.

The explicit forms of the thermal spin current and the spin current flux have been derived for the degenerate electrons. Corresponding modification of the quantum vorticity evolution equation has been found. The thermal part of the spin current has been applied to find the spectrums of the transverse waves, focusing on the spin-plasma waves, in the electron-ion and electron-positron plasmas. It was demonstrated that the thermal spin current decreases the frequencies of the spin-plasma waves with right circular polarization propagating parallel and perpendicular to the external magnetic field. We have found that for large enough TPSC the spin-plasma waves with right circular polarization propagating parallel to the external magnetic field disappears, but the right circularly polarized spin-plasma wave arises at frequency near the TSCF. If the TSCF and the Langmuir frequency are approximately equal to each other we have obtained hybridization of the spin-plasma wave and the ordinary electromagnetic wave and their spectrums.

All of it have been found as applications of the generalized non-linear Pauli equation with the spinor pressure term suggested in this paper.

\begin{acknowledgements}
The work of P.A. was supported by the Dynasty foundation.
\end{acknowledgements}

\end{document}